\documentclass[times,12pt]{article}

\usepackage{times}
\usepackage{balance}
\usepackage{graphicx}
\usepackage{amsmath, amsthm}
\usepackage{amssymb}
\usepackage{setspace}
\usepackage[margin=1in]{geometry}
\usepackage{subfigure}

\RequirePackage[colorlinks,citecolor=blue,urlcolor=blue]{hyperref}
\usepackage{breakurl}

\def\argmin{\mathop{\rm argmin}}

\usepackage{natbib,upgreek}

\usepackage{bbm}
\usepackage{multirow}
\usepackage{doi}

\begin{document}

\title{Regression-based covariance functions for nonstationary spatial modeling}

\author{Mark D. Risser$^{\> a}$ and Catherine A. Calder$^{\>a \>*}$}
\maketitle

{\scriptsize
\hskip4ex $^a$ Department of Statistics, The Ohio State University, OH 43210, USA

\hskip4ex $^*$ Correspondance: Department of Statistics, 1958 Neil Avenue, Columbus, OH 43210. Email: \tt{calder@stat.osu.edu}
}

\doublespacing

\begin{abstract}

In many environmental applications involving spatially-referenced data, limitations on the number and locations of observations motivate the need for practical and efficient models for spatial interpolation, or kriging. A key component of models for continuously-indexed spatial data is the covariance function, which is traditionally assumed to belong to a parametric class of stationary models. However, stationarity is rarely a realistic assumption. Alternative methods which more appropriately model the nonstationarity present in environmental processes often involve high-dimensional parameter spaces, which lead to difficulties in model fitting and interpretability. To overcome this issue, we build on the growing literature of covariate-driven nonstationary spatial modeling. Using process convolution techniques, we propose a Bayesian model for  continuously-indexed spatial data based on a flexible parametric covariance regression structure for a convolution-kernel covariance matrix. The resulting model is a parsimonious representation of the kernel process, and we explore properties of the implied model, including a description of the resulting nonstationary covariance function and the interpretational benefits in the kernel parameters. Furthermore, we demonstrate that our model provides a practical compromise between stationary and highly parameterized nonstationary spatial covariance functions that do not perform well in practice. We illustrate our approach through an analysis of annual precipitation data.

\end{abstract}


\section{Introduction}
\label{sec:intro}

In spite of the rising popularity of spatio-temporal modeling, there is still a strong need for flexible spatial models appropriate for spatial prediction. For example, in the case of meteorological, agricultural, or geological data where fixed monitoring stations are used to collect observations of a spatial process, monitoring sites are not always located where information about the spatial process is desired. Alternatively, it may be of interest to generate a ``filled-in'' prediction map of the spatial process based on a sparse, finite number of observations, as well as estimate the uncertainty in these predictions. A Gaussian process (GP) is a popular way to model environmental processes in order to answer questions such as these, and an important component of a GP model is the covariance function, which describes the degree and nature of spatial dependence present in a spatial process. It should be noted that the covariance function is also a key component of geostatistical or kriging methods, which do not make the Gaussian process assumption. In either case, parametric models for a spatial covariance function typically require the strong assumption of stationarity, in which the spatial dependence between two locations is a function of only their separation vector or distance, if isotropy is assumed. This modeling assumption is made mostly for convenience and is rarely appropriate in real-world applications. 

As a result, a variety of alternative methodologies for modeling second-order nonstationarity in spatial processes have been developed, most prominently deformation techniques (\citealp{sampGutt}, \citealp{damian2001}, \citealp{schmidt03}, \citealp{anderes2008}), basis function expansions (\citealp{holland},  \citealp{nychka2002}, \citealp{PintoreHolmes}, \citealp{matsuo2011}, \citealp{katzfuss2013}), Markov random field models using stochastic partial differential equations (SPDEs) (\citealp{Lindgren2011}), and process convolution (PC) methods. Process convolutions (also called kernel smoothing or moving average models) provide a constructive approach to specifying a nonstationary GP. In a PC model, a continuous latent process is convolved with a smoothing kernel function, where either the process or kernel vary over space. \cite{Higdon98} and \cite{Higdon99} opt to fix the latent process and specify a spatially-varying kernel function (see also \citealp{PacScher}). Alternatively, \cite{fuentes02} instead fixes the kernel function and allows the latent process to be spatially dependent (see also \citealp{fuentes2001}). 

Building off the intuition of mean regression, more recent methodology uses the idea that covariate information might play a useful role in specifying the covariance structure of a spatial process. By covariate information we mean spatially-varying, observable quantities which can either be collected at all prediction locations of interest or in some way interpolated from nearby observations (for example, elevation, wind speed or direction, soil quality, proximity to a pollution source or geographical feature, etc.). In general, the argument for using covariate information in a covariance function is both interpretational and computational. First, the major drawback to not using covariate information to model spatial dependence is that it becomes difficult to understand why the process exhibits nonstationary behavior, i.e., how the dependence structure changes over space. Introducing covariates in a covariance function is a natural way to impose the desired second-order nonstationarity and allows for an explanation of the spatially-varying dependence structure. Secondly, many of the ``non-covariate'' approaches are highly parameterized and therefore difficult to implement, since it is hard to estimate nonstationary behavior using only a single realization of a spatial process. As in mean regression, using covariates allows the dimension of the requisite parameter space to be greatly reduced, facilitating computation.

Indeed, some work has been done to expand the aforementioned nonstationary approaches to incorporate covariate information. \cite{schmidt11} use covariate information in a deformation model, and a recent extension of the \cite{Lindgren2011} work that includes covariates is given in \cite{spde13}. Covariate information was first introduced in the \cite{Higdon98} version of a  process convolution model by \cite{calder08}, who, following \cite{Higdon98}, chose the kernel functions to be Gaussian but used covariate information to fix the kernel function parameters. Alternatively, \cite{ViannaNeto} introduce a convolution approach that also incorporates directional covariates, and \cite{reich2011} propose a spatio-temporal model with covariate information for the \cite{fuentes02} kernel-smoothing method.

While all of these approaches have been shown to be successful, limitations remain. Several methods are only appropriate for directional covariates (e.g., \citealp{ViannaNeto}) while others are not fully Bayesian (e.g., \citealp{calder08}). More seriously, while successfully including covariate information, many of the methods fail to address the issue of characterizing \textit{how} a spatially-varying covariate impacts the covariance function. For example, \cite{reich2011} use covariate information to model the weights for each of the stationary spatial processes, not the spatial dependence properties of these processes. 

In this work, we address these limitations by proposing new methodology which is fully Bayesian and allows covariate information to be included directly in a model for the spatial dependence properties of the resulting covariance function. Furthermore, the model is able to accommodate any type of covariate information, be it scalar or directional, discrete or continuous, and yields a parsimonious parameterization so that a relatively fast, stable, and efficient model fitting algorithm can be implemented. Finally, the parameters allow for interpretations of how the covariate impacts the spatial dependence, and, given parameter estimates, the changes in spatial dependence over the region of interest are easily visualized. The resulting model is applicable in any geostatistical setting in which a spatial Gaussian process model is appropriate; e.g., modeling and prediction of environmental, meteorological, and pollution- or disease-related processes. 

The paper is structured as follows. In Section \ref{sec:cov_fcn}, we present the motivation for and derivation of our proposed covariance function, as well as the idea of parametric covariance regression for the kernel parameters. The implied properties of the covariance function are explored, including model geometry and interpretational benefits in the kernel parameters. We also show that stationary and isotropic models are special cases of this model. In Section \ref{modelstatement} we outline a fully Bayesian model, which is applied in Section \ref{application} to a real world example. Commentary and thoughts for continued research are provided in Section \ref{discussion}.

\section{A class of regression-based nonstationary covariance functions} \label{sec:cov_fcn}

\subsection{The nonstationary Mat\'{e}rn class}
{\sloppypar
A mean zero spatial Gaussian process $Y(\cdot)$ on $G \subset \mathcal{R}^d$ can be defined by the kernel convolution 
\begin{equation*}
Y({\bf s}) = \int_{G} K_{\bf s}({\bf u}; \boldsymbol{\phi}) dW({\bf u}),
\end{equation*}
where $W(\cdot)$ is $d$-dimensional Brownian motion and $K_{\bf s}(\cdot; \boldsymbol{\phi})$ is a spatially-varying parametric kernel function centered at ${\bf s} \in G$. The requirements on the kernel function are simply that $\int_{\mathcal{R}^d} K_{\bf s}({\bf u}; \boldsymbol{\phi}) d{\bf u} <\infty$ and $\int_{\mathcal{R}^d} K^2_{\bf s}({\bf u}; \boldsymbol{\phi}) d{\bf u} <\infty$. In general, the covariance function corresponding to this process is $C({\bf s}_i, {\bf s}_j; \boldsymbol{\phi}) = E \big[Y({\bf s}_i)Y({\bf s}_j) \big] = \int_{G}  K_{{\bf s}_i}({\bf u}; \boldsymbol{\phi}) K_{{\bf s}_j}({\bf u}; \boldsymbol{\phi}) d{\bf u}$ where ${\bf s}_i, {\bf s}_j \in G$; the benefit of this constructive approach lies in the fact that it is much easier to specify kernel functions than a covariance function directly. If $d$-variate Gaussian densities are used for the kernel functions (\citealp{thiebaux76}; \citealp{thiebaux_pedder}), this integral can be calculated analytically and a closed form can be obtained for the covariance function, namely 
\begin{equation} \label{cov}
C({\bf s}_i, {\bf s}_j; \boldsymbol{\phi}) = (2\pi)^{-\frac{d}{2}} \left|\frac{{\bf \Sigma}({\bf s}_i) + {\bf \Sigma}({\bf s}_j)}{2}\right|^{-\frac{1}{2}} \exp \{-Q_{ij} \}
   =  (2\pi)^{-\frac{d}{2}} \left|\frac{{\bf \Sigma}({\bf s}_i) + {\bf \Sigma}({\bf s}_j)}{2}\right|^{-\frac{1}{2}} \mathcal{G}\left(\sqrt{Q_{ij}}\right),
\end{equation}
where ${\bf \Sigma(s)}$ is the $d \times d$ covariance matrix for the Gaussian kernel function centered at location ${\bf s}$ (henceforth called the kernel matrix), 
\begin{equation*} \label{Qij}
Q_{ij} =  ({\bf s}_i-{\bf s}_j)' \left(\frac{{\bf \Sigma}({\bf s}_i) + {\bf \Sigma}({\bf s}_j)}{2}\right)^{-1}({\bf s}_i-{\bf s}_j)
\end{equation*}
is a scaled squared separation length, and $\mathcal{G}(\cdot)$ is the standard Gaussian correlation function. A full derivation of (\ref{cov}) is given in the appendices of \cite{PacScher}. However, as discussed in \cite{PacScher}, using a Gaussian correlation function as in (\ref{cov}) has the undesirable property of giving process realizations which are infinitely differentiable and therefore too smooth for most applications. Building off the ideas in \cite{paciorek2003}, \cite{stein2005} proves that a generalization of (\ref{cov}) still gives a valid covariance function. Specifically,
\begin{equation} \label{stein_cov}
C^{NS}({\bf s}_i, {\bf s}_j; \boldsymbol{\phi}) = \sigma({\bf s}_i) \sigma({\bf s}_j) \left|\frac{{\bf \Sigma}({\bf s}_i) + {\bf \Sigma}({\bf s}_j)}{2}\right|^{-1/2} \mathcal{M}_{\frac{\nu({\bf s}_i) + \nu({\bf s}_j)}{2}}\left(\sqrt{Q_{ij}} \right)
\end{equation}
is a valid (nonstationary) covariance function, where $\sigma(\cdot)$ is a spatially-varying ``standard deviation'' process ($\sigma^2({\bf s})$ is proportional to the process variance at ${\bf s}$), $\nu(\cdot)$ is a spatially-varying smoothness process, ${\bf \Sigma(\cdot)}$ is a kernel matrix process, and $\mathcal{M}_\nu(\cdot)$ is the Mat\'{e}rn correlation function with smoothness $\nu$. The kernel matrix process can be interpreted as a spatially-varying geometric anisotropy process; i.e., ${\bf \Sigma(s)}$ controls the anisotropic behavior of the process $Y(\cdot)$ in a small neighborhood of ${\bf s}$. The spatial covariance model in (\ref{stein_cov}) is extremely flexible, as it allows the variance, smoothness, and geometric anisotropy of $Y(\cdot)$ to vary over space while maintaining a closed form. Furthermore, using the Mat\'{e}rn class of correlation functions avoids the undesirable smoothness properties of (\ref{cov}). Note that while the Mat\'{e}rn correlation function is used here, the resulting covariance function will still be positive definite and valid on $\mathcal{R}^d$, for $d = 1, 2, \ldots$, when any other isotropic correlation function which is valid on $\mathcal{R}^d, d=1, 2, \ldots$ is used in place of $\mathcal{M}_{\nu}\left(\cdot \right)$ in (\ref{regression_covfcn}), such as the spherical correlation function or even a compactly supported correlation function (\citealp{PacScher}).

The covariance function (\ref{stein_cov}) has been used in various forms throughout the literature.  \cite{PacScher} fix $\nu({\bf s}) \equiv \nu$ ($\nu$ an unknown constant) for all ${\bf s}$ and take $\sigma({\bf s}) = \sigma  \left|{\bf \Sigma(s)} \right|^{1/4}$ ($\sigma$ an unknown constant), which results in a constant process variance over the spatial region.  \cite{Anderes_Stein} find that it is difficult to separate the effect of ${\bf \Sigma(s)}$ and $\nu({\bf s})$ if (\ref{stein_cov}) is used directly; instead, they constrain the kernel matrices to be a multiple of the identity matrix and a introduce a separate model for $\nu(\cdot)$. \cite{kleiber2012} use (\ref{stein_cov}) directly for multivariate spatial processes. None of these approaches incorporate covariate information. 

A major difference in our approach is that we will model $\sigma(\cdot)$, $\nu(\cdot)$, and ${\bf \Sigma}(\cdot)$ as parametric functions instead of stochastic processes. As a result, the parameters will still be spatially-varying, but the dimension of the resulting parameter space will be greatly reduced. 
As discussed in the introduction, the case for including covariate information in a model for the second-order properties of a spatial process is strong, and therefore the spatially-varying nature of these models will be driven by covariate information. Specifically, we assume that for each ${\bf s} \in G$ we have ${\bf x(s)} = (1, x_1({\bf s}), ... , x_{p-1}({\bf s}))' \in \mathcal{R}^{p}$, a vector of observable, spatially-varying covariate information (including an intercept).

Since the standard deviation function $\sigma(\cdot)$ is scalar and simply needs to be positive, a simple log-linear regression model can be introduced, namely
\begin{equation} \label{var_reg}
\sigma({\bf s}) = \exp \left\{ \frac{\boldsymbol{\alpha} \hskip0.25ex {\bf x(s)}}{2} \right\}  \left|{\bf \Sigma(s)} \right|^{1/4},
\end{equation}
where $\boldsymbol{\alpha}$ is a vector of coefficients that control the impact of the covariate on $\sigma(\cdot)$. Including $\left|{\bf \Sigma(s)} \right|^{1/4}$ in (\ref{var_reg}) separates the effect of the variance and kernel functions; the resulting variance is 
$
Var \big(Y({\bf s}) \big) = \exp\{ \boldsymbol{\alpha} \hskip0.25ex {\bf x(s)} \}.
$
Therefore, the elements of $\boldsymbol{\alpha}$ have the usual log-linear regression interpretations with respect to the covariates. Similarly, the smoothness function can also be modeled as a log-linear function of covariates, for example,
\begin{equation} \label{smooth_reg}
\nu({\bf s}) = \exp \left\{ \frac{\boldsymbol{\delta} \hskip0.25ex {\bf x(s)}}{2} \right\}.
\end{equation}
As with $\boldsymbol{\alpha}$, the coefficients in $\boldsymbol{\delta}$ have straightforward interpretations.

The parametric model for ${\bf \Sigma(s)}$ uses the idea of covariance regression from \cite{hoff_niu}, in which the kernel matrices are parameterized as 
\begin{equation} \label{kern_reg}
{\bf \Sigma}{\bf (s)} = {\bf \Psi} + {\bf \Gamma x(s) x(s)'} {\bf \Gamma}'.
\end{equation}
In the original paper, this parametric form is a consequence of a random-effects representation for multivariate outcomes, where ${\bf \Psi}$ represents an error covariance and the coefficients of ${\bf \Gamma}$ describe how additional variability is distributed across the $d$ dimensions. The constraints on the parameters in this model simply require ${\bf \Psi}$ to be a $d \times d$ positive definite, symmetric matrix while ${\bf \Gamma}$ can be any $d \times p$ real matrix. In this model, ${\bf \Psi}$ is identifiable and ${\bf \Gamma}$ is identifiable up to a sign, given an adequate range of $x$-values (\citealp{hoff_niu}). 

Specifying the process variance as in (\ref{var_reg}) and the kernel matrices according to (\ref{kern_reg}) gives rise to a new class of ``regression'' spatial covariance functions, which are a special case of (\ref{stein_cov}) but allow for incorporation of covariate information. Defining ${\bf x}_i \equiv {\bf x}({\bf s}_i)$ and ${\bf x}_j \equiv {\bf x}({\bf s}_j)$, the covariance function is
\begin{equation} \label{regression_covfcn} 
C^{R}({\bf s}_i, {\bf s}_j; \boldsymbol{\alpha}, \boldsymbol{\delta}, {\bf \Psi, \Gamma}) = \exp \left\{  \boldsymbol{\alpha}\left(\frac{{\bf x}_i + {\bf x}_j}{2} \right) \right\} \frac{|{\bf \Psi + \Gamma }{\bf x}_i {\bf x}_i' {\bf \Gamma}'|^{\frac{1}{4}}  |{\bf \Psi + \Gamma}{\bf x}_j {\bf x}_j' {\bf \Gamma}'|^{\frac{1}{4}}}{ \left|{ \bf \Psi + \Gamma}\left(\frac{{\bf x}_i{\bf x}'_i + {\bf x}_j{\bf x}'_j}{2} \right){\bf \Gamma'}\right|^{\frac{1}{2}}} \mathcal{M}_{ \exp \left\{\boldsymbol{\delta}\left(\frac{{\bf x}_i + {\bf x}_j}{2}\right) \right\}}\left(\sqrt{Q_{ij}} \right),
\end{equation}
where again 
\begin{equation*}
Q_{ij} =  ({\bf s}_i-{\bf s}_j)' \left({ \bf \Psi + \Gamma}\left(\frac{{\bf x}_i{\bf x}'_i + {\bf x}_j{\bf x}'_j}{2} \right){\bf \Gamma'}\right)^{-1}({\bf s}_i-{\bf s}_j).
\end{equation*}
Note that $C^{R}$ is still a nonstationary covariance function, but none of the parameters vary spatially. The ``$R$'' superscript indicates that the kernel matrices, variance, and smoothness will be modeled in a regression framework using covariate information. This covariance function is valid on $\mathcal{R}^d$, $d\geq1$; the proof of this result is a direct corollary of the more general proof for the validity of (\ref{stein_cov}) (given in \citealp{stein2005}). 
}
\subsection{Parameter interpretations, model geometry, and parsimony} \label{Kern_Reg}

To illustrate how the parameters in this model are related to the spatial dependence properties of $Y(\cdot)$, we first consider a one-dimensional example ($d=1$). Here, we suppose the process of interest defined on $G \subset \mathcal{R}^1$ and explore the properties of ${\bf \Sigma(\cdot)}$ as a function of the covariate ${\bf x}(s) = \left( 1, x(s) \right)'$, consisting of an intercept and a continuous and differentiable covariate $x(s)$  (so that $p=2$). The covariance regression parameters are then ${\bf \Psi} = \psi \in \mathcal{R^+}$ and ${\bf \Gamma} = \boldsymbol{\gamma}' \in \mathcal{R}^{1\times 2}$, and the model defines kernel variances (instead of matrices) for each location. For a generic location $s$, the kernel variance function is
\begin{equation} \label{1dkernvar}
\Sigma( s ) = \psi + {\bf  \boldsymbol{\gamma}' x}(s) {\bf x}(s)' \boldsymbol{\gamma} = \psi + (\gamma_{11} + \gamma_{12} x )^2,
\end{equation}
where we suppress the dependence of $x$ on location $s$ in the right hand side of (\ref{1dkernvar}), as we view $\Sigma(s)$ as a function of $x$ in the following discussion. In (\ref{1dkernvar}), for fixed values of the parameters $\psi, \gamma_{11}$, and $\gamma_{12}$, we see the kernel variance $\Sigma(s)$ is a smooth, convex, quadratic function of the covariate $x$. For fixed $\psi, \gamma_{11}$, and $\gamma_{12}$, the minimum of $\Sigma(s)$ is $\psi$, attained at $\hat{x} = -\gamma_{11}/\gamma_{12}$. Thus, we can interpret $\psi$ as the minimum kernel variance allowed in the model, attained for locations where the covariate $x$ is equal to $-\gamma_{11}/\gamma_{12}$. In a spatial context, ``minimum kernel variance'' is equivalent to smallest correlation between points a fixed ``distance'' apart, where ``distance'' is described below. The derivative of the kernel variance is $2\gamma_{12}^2 x + 2\gamma_{11}\gamma_{12}$; we can further interpret $|\gamma_{12}|$ as controlling the rate of change in $\Sigma(s)$ with respect to $x$ (as well as indicating the degree of nonstationarity in the kernel variance). 

In this simple case, the ``distance'' measure is 
\begin{equation*}
\sqrt{Q_{ij}} = \frac{|s_i - s_j|}{\phi(s_i, s_j)},
\end{equation*}
where $\phi(s_i, s_j) =  \sqrt{\big(\Sigma(s_i) + \Sigma(s_j)\big)/2}$. Note that ${Q_{ij}}$ is not a true distance measure, as it violates the triangle inequality. Regardless, the model (\ref{kern_reg}) for the kernel variances also dictates the range of the spatial process $Y(\cdot)$, scaling distances according to the covariate and parameter values. Explicitly, the squared range function is 
\begin{equation} \label{1dphi}
\phi^2(s_i, s_j) \equiv \frac{\Sigma(s_i) + \Sigma(s_j)}{2} = \gamma_{12}^2\left( \frac{x_i^2 + x_j^2}{2} \right) + 2 \gamma_{11}\gamma_{12} \left( \frac{x_i + x_j}{2} \right) + (\gamma_{11}^2 + \psi).
\end{equation}
Again this function is quadratic in each of its inputs, although there is also an interesting averaging property between the covariates at two locations:  $\phi^2(s_i, s_j)$ depends on both $x_i^2 + x_j^2$ and $x_i + x_j$. This functional form in (\ref{1dphi}) hints at further interpretations for the parameters in this simple model. Like $\Sigma(s)$, $\phi$ is a smooth, convex function of the covariates. Furthermore,
\begin{equation*}
\argmin_{s, s' \in G} \phi(s,s') = \left(-\frac{\gamma_{11}}{\gamma_{12}}, -\frac{\gamma_{11}}{\gamma_{12}} \right), \hskip2ex \min_{s, s' \in G} \phi(s,s') = \sqrt{\psi},
\end{equation*}
and again the rate of change in $\phi$ is controlled by $|\gamma_{12}|$. Thus, we can also interpret $\psi$ to correspond to the square of the minimum spatial range in the covariance function, and pairs of locations having minimum spatial correlation (those with a covariate value $-\gamma_{11}/\gamma_{12}$) also minimize the spatial range function, which is an intuitive property. Note that in one dimension, the range is a heuristic representation for the ``width'' of the kernel function with variance ${ \Sigma(\cdot)}$, as opposed to the more conventional geostatistical definition of range, which corresponds to the distance at which the semivariogram reaches 95 percent of the sill.

Returning to the more realistic two-dimensional case (i.e., $d=2$) but for now keeping $p=2$, so that the covariate vector is still ${\bf x(s)}' = \left( 1, x \right)$, the kernel matrix function becomes 
\begin{equation} \label{2dkernmat}
\begin{array}{rcl}
{\bf \Sigma(s)} & = & {\bf \Psi + \Gamma x(s) x(s)' \Gamma'} = \left[ \begin{array}{cc}
\psi_{11} + {\bf \boldsymbol{\gamma}_1' x(s) x(s)' \boldsymbol{\gamma_1}} & \psi_{12} + {\bf \boldsymbol{\gamma}_1' x(s) x(s)' \boldsymbol{\gamma}_2} \\
\psi_{12} + {\bf \boldsymbol{\gamma}_2' x(s) x(s)' \boldsymbol{\gamma}_1} & \psi_{22} + {\bf \boldsymbol{\gamma}_2' x(s) x(s)' \boldsymbol{\gamma}_2}  \\
\end{array} \right] \\
 & \equiv &  \left[ \begin{array}{cc} \Sigma_{11}({\bf s}) & \Sigma_{12}({\bf s }) \\ \Sigma_{12}({\bf s}) & \Sigma_{22}({\bf s}) \end{array} \right],
\end{array}
\end{equation}
where $\boldsymbol{\gamma}_1'$ and $\boldsymbol{\gamma}_2'$ are the row vectors of ${\bf \Gamma}$, and $\Sigma_{ij}( {\bf s}) = \psi_{ij} + \gamma_{i1}\gamma_{j1} + (\gamma_{i1}\gamma_{j2} + \gamma_{i2}\gamma_{j1})x + \gamma_{i2}\gamma_{j2}x^2$, for $i, j = 1, 2$. Because the $2\times 2$ kernel matrix is parameterized directly and not in terms of its spectral decomposition, it becomes more difficult to interpret these parameters in the traditional ways (i.e., in terms of the magnitude and direction of spatial range). However, several general comments can be made. First, because the diagonal elements of ${\bf \Gamma x(s) x(s)' \Gamma'}$ are necessarily non-negative, ${\bf \Psi}$ represents a ``baseline'' or minimum kernel matrix. The marginal kernel variances (i.e., $\Sigma_{11}({\bf s})$ and $\Sigma_{22}({\bf s})$) can be interpreted in exactly the same way as the one dimensional case (note the similarity to the one-dimensional kernel variance in (\ref{1dkernvar})). However, aside from ${\bf \Psi}$, it is important to notice that the off-diagonal element is modeled with the same parameters as the marginal variances. That is, because the magnitude of
\begin{equation*}
\Sigma_{12}({\bf s}) = \psi_{12} + (\gamma_{11} + \gamma_{12}x)(\gamma_{21} + \gamma_{22}x)
\end{equation*}
increases as $(\gamma_{11} + \gamma_{12}x)$ and $(\gamma_{21} + \gamma_{22}x)$ increase (which increases the marginal variances), these will increase the magnitude of the covariance as well. Finally, because the squared ``distance'' measure $Q_{ij}$ involves
\begin{equation} \label{2davg}
\frac{{\bf \Sigma}({\bf s}_i) + {\bf \Sigma}({\bf s}_j)}{2} = \frac{1}{2} \left[ \begin{array}{cc}
\Sigma_{11}( {\bf s}_i) + \Sigma_{11}( {\bf s}_j) & \Sigma_{12}( {\bf s}_i) + \Sigma_{12}( {\bf s}_j) \\
\Sigma_{12}( {\bf s}_i) + \Sigma_{12}( {\bf s}_j) & \Sigma_{22}( {\bf s}_i) + \Sigma_{22}( {\bf s}_j) \\
\end{array} \right],
\end{equation}
the covariance regression model for the kernel matrices (\ref{kern_reg}), which by nature models geometric anisotropy, again implies a parametric structure for the range of the process. (Note the similarity of the diagonal elements of (\ref{2davg}) to (\ref{1dphi}).) Intuitively, for $d=2$, the spatial range is again a representation for the ``width'' or volume of the anisotropy ellipse corresponding to ${\bf \Sigma(\cdot)}$.

\begin{figure}[h]
\includegraphics[width=\textwidth]{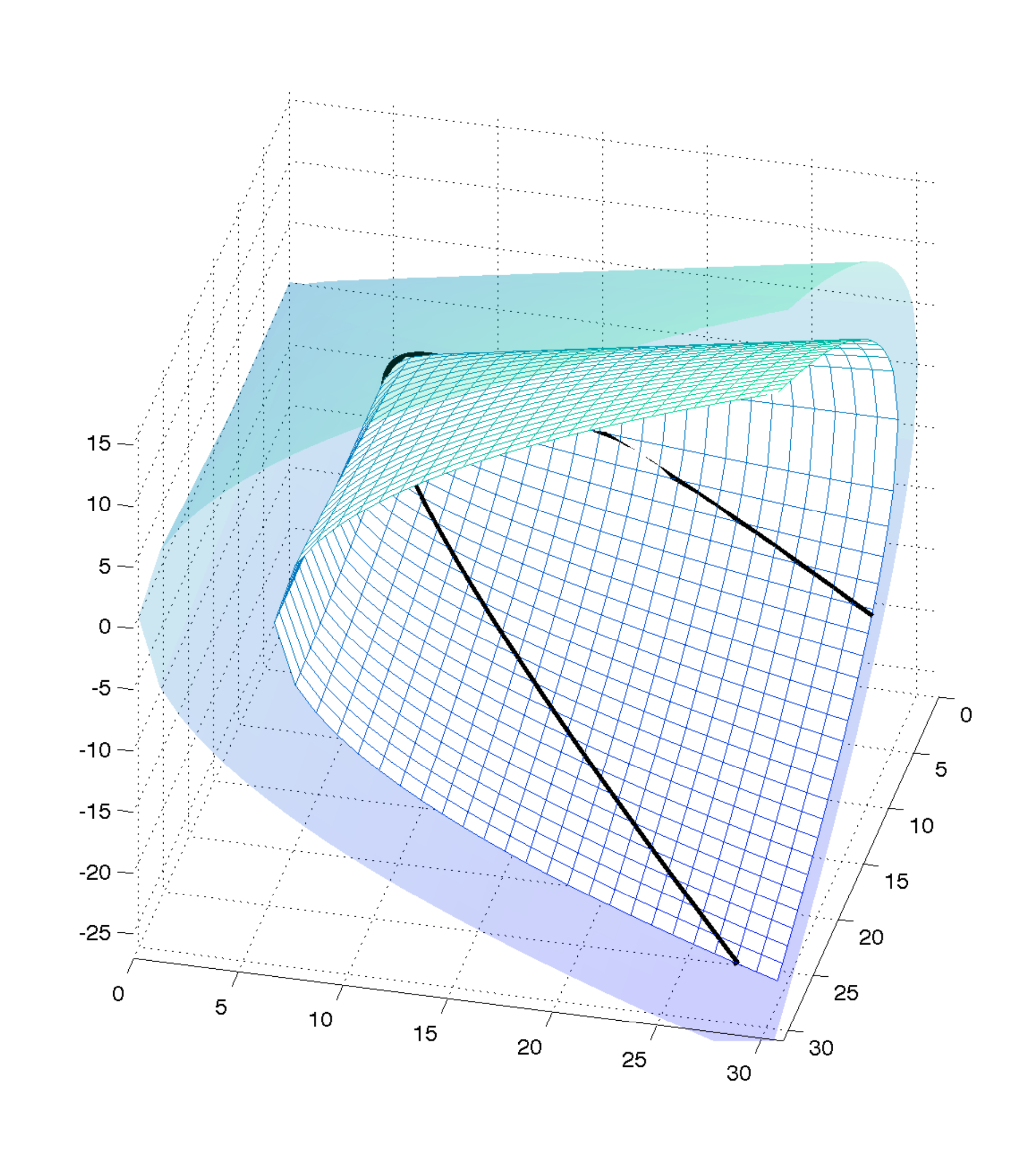}
\caption{ A visualization of the parameter space for two-dimensional kernel matrices under the covariance regression model. }
\label{paramspace}
\end{figure}

The parametric model (\ref{kern_reg}) imposes a particular model geometry which is a restriction on the unconstrained parameter space of positive definite matrices. Generically denote a $2\times 2$ kernel (covariance) matrix as
\begin{equation*}
{\bf \Sigma} = \left[ \begin{array}{cc} \sigma_1^2 & \sigma_{12} \\ \sigma_{12} & \sigma_2^2 \end{array} \right],
\end{equation*}
where positive-definiteness requires $\sigma_1 > 0$, $\sigma_2 > 0$, and $|\sigma_{12}| < \sigma_1 \sigma_2$. Using these constraints, the parameter space in two dimensions is a three-dimensional cone (shown in Figure \ref{paramspace} by the opaque cone) where the $x$-coordinate represents $\sigma_1^2$, the $y$-coordinate represents $\sigma_2^2$, and the $z$-coordinate represents $\sigma_{12}$. The covariance regression model (\ref{kern_reg}) restricts the parameter space to give kernel matrices that can be represented by three-dimensional coordinates
$
(x,y,z) = \big(  \Sigma_{11}({\bf s}), \Sigma_{22}({\bf s}), \Sigma_{12}({\bf s}) \big)
$
where $\Sigma_{ij}( {\bf s}) = \psi_{ij} + (\gamma_{i1} + \gamma_{i2}x)(\gamma_{j1} + \gamma_{j2}x)$. A fixed value of ${\bf \Psi}$ has the effect of ``pushing'' the cone inside the large (full parameter space) cone by orienting the tip of the recessed cone at $(\psi_{11}, \psi_{22}, \psi_{12})$. For a fixed choice of ${\bf \Gamma}$, the component ${\bf \Gamma x x' \Gamma'}$ is a quadratic function of ${\bf x}$ but also a rank 1 matrix (${\bf x}$ is a vector). Therefore, ${\bf \Gamma}$ further restricts the parameter space
to fall within a curve on the surface of the recessed cone. Both of these restrictions are shown in Figure  \ref{paramspace} along with the full space. The important thing to note in this plot is that the kernel matrix function (\ref{2dkernmat}) is still a smooth, convex function of the covariate ${\bf x}$, and the spatially-varying geometric anisotropy of the spatial process will change smoothly according to covariate information when parameterized according to (\ref{kern_reg}).

The regression parameterization of a nonstationary spatial covariance function is extremely parsimonious, a consequence of allowing the spatially-varying properties to depend completely upon the covariate. The dimension of the parameter vector to be estimated for the kernel matrix function is
$
dim({\bf \Psi, \Gamma}) = d(d+1)/2 + dp,
$
and it is important to note that in this representation, the dimension of the parameter space does not increase with $n$, the number of observations. Furthermore, in a typical spatial application where $d=2$, if only a single covariate is used (i.e., $p=2$), there are just seven parameters to estimate.

This is in stark contrast to the model fit in \cite{PacScher}, which, while extremely flexible, is notoriously difficult to fit (see, e.g., \citealp{ViannaNeto}). This problem arises because it is desirable for a nonstationary covariance function to have locally stationary properties, which will be the case if the kernel matrices vary smoothly over space (\citealp{PacScher}). In Paciorek's paper, this regularization of the kernel matrices is accomplished by assigning the elements of these matrices Gaussian process priors, and then fitting the model with a nonparametric basis function approximation to the Gaussian process. However, it is not possible to marginalize over these kernel matrices, so the dimension of the parameter space is linear in the sample size.

\subsection{General properties of $C^R$} \label{sec:properties}

Regardless of the spatial dimension $d$ or covariate dimension $p$, general statements can be made with respect to the properties of the regression covariance function $C^R$ and why (\ref{regression_covfcn}) might prove to be useful in theory and in practice.

First, the covariance function (\ref{regression_covfcn}) is nonstationary in both geographic and covariate space. Specifically, $C^R$ is a function of
\begin{equation*}
\frac{ {\bf \Sigma}({\bf s}_i) + {\bf \Sigma}({\bf s}_j)}{2} = { \bf \Psi + \Gamma}\left(\frac{{\bf x}_i{\bf x}_i' + {\bf x}_j{\bf x}_j'}{2} \right){\bf \Gamma'},
\end{equation*}
which cannot be expressed in terms of $||{\bf x}_i - {\bf x}_j||$ or even $({\bf x}_i - {\bf x}_j)$.  This is a property similar to the model in \cite{reich2011}, and is desirable. For example, consider pairs of points $({\bf s}_i, {\bf s}_j)$ and $({\bf s}_i', {\bf s}_j')$ such that both ${\bf x}({\bf s}_i) - {\bf x}({\bf s}_j) = {\bf x}({\bf s}_i') - {\bf x}({\bf s}_j')$ and ${\bf s}_i - {\bf s}_j = {\bf s}_i' - {\bf s}_j'$. 
Despite the fact that these two pairs of points have the same difference in their covariate value and the same spatial lag, the model can specify different spatial dependencies for each pair. Intuitively, a collection of monitoring sites with the same high elevation could display a different dependence pattern than another collection of monitoring sites with the same low elevation, and it is important to be able to capture this difference. 

Secondly, using (\ref{regression_covfcn}) results in intuitive local properties for the underlying spatial process. Because the kernel matrices and process variance are modeled according to covariate information, the covariate dictates the implied behavior of the spatial process. If the covariate of interest varies smoothly, for locations in a small neighborhood of a point $ {\bf s_0}$ ($\{{\bf s: ||s-s_0||} \leq \delta\}$, $\delta$ small) the kernel matrices and process variance will be essentially constant (because ${\bf x(s) \approx x(s_0)}$), and
\begin{equation*}
C^{R}({\bf s, s_0}; \boldsymbol{\alpha}, \boldsymbol{\delta}, {\bf \Psi, \Gamma}) \approx \exp \{ \boldsymbol{\alpha}{\bf x_0} \} \mathcal{M}_{ \exp \left\{\boldsymbol{\delta}{\bf x_0}\right\}}\left(\sqrt{{\bf (s-s_0)}' {\bf \Sigma(s_0)}^{-1}{\bf (s-s_0)}} \hskip1ex \right).
\end{equation*}
Because this depends only on $({\bf s-s_0})$, this covariance function is again stationary (anisotropic), while still allowing the covariate to dictate the spatial dependence properties. As an aside, it should be noted that if, on the other hand, the covariate does not vary smoothly (e.g., for a discrete or categorical covariate), the model still implies stationary behavior for regions or groups of locations that share a common covariate. That is, for any region $A \subset G$ such that ${\bf x(s) = x_0}$ for all ${\bf s} \in A$, the regression covariance function is exactly
\begin{equation*}
C^{R}({\bf s}_i, {\bf s}_j; \boldsymbol{\alpha}, \boldsymbol{\delta}, {\bf \Psi, \Gamma}) = \exp \{ \boldsymbol{\alpha}{\bf x_0} \}   \mathcal{M}_{ \exp \left\{\boldsymbol{\delta}{\bf x_0}\right\}}\left(\sqrt{ ({\bf s}_i - {\bf s}_j)' {\bf \Sigma(x_0)}^{-1}({\bf s}_i - {\bf s}_j) }\hskip1ex \right)
\end{equation*}
for any ${\bf s}_i, {\bf s}_j \in A$. However, the model still allows for non-zero covariance between two points that are not in the same region and do not have a common value of the covariate. In this way, the model mimics that of \cite{fuentes02}, who obtains a nonstationary model by averaging processes that are stationary over distinct regions.

However, this is not the only way in which the property of stationarity can be recovered. Another benefit of using a regression framework for the kernel matrices and process variances is that a stationary model is a special case of the more general nonstationary model (\ref{regression_covfcn}), which is true when certain components of ${\bf \Gamma}$ are zero. Thus, the data can indicate a lack of nonstationarity in a spatial process, regardless of the modeling assumptions. The array of other tools available in mean regression are useful here as well: for example, if multiple covariates are to be used in the covariance function it can be determined which covariates are important in determining nonstationarity.

Additionally, the covariance function can accommodate any type of meaningful covariate information. This flexibility is in fact what qualifies this model for use in a wide variety of geostatistical settings, since different types of spatial processes are likely to be influenced by different types of covariate information. For example, weather-related processes (such as annual precipitation or average temperature) might relate closely to topographical features (such as elevation), while  airborne processes (vector-borne diseases, radiation, or pollution) might be heavily affected by known geographical features and directional wind patterns. More generally, the covariate can even be a categorical or discrete variable, and if a vector-valued covariate is of interest, there is no need to derive a scalar representation  (as in \citealp{ViannaNeto}) -- the multi-dimensional characteristics of the covariate can be entered into the model individually. Furthermore, as a regression model, many model selection techniques developed for mean regression can be applied here, with the possibility of including interaction terms or higher-order functions of a covariate. 

Finally, recall that a major goal for this model is to be able to give easily obtainable summaries of how dependence properties change in space according to covariate information. Given estimated values of the parameters, we can easily provide a visual summary of important second-order properties, illustrating how dependence changes over space. For example, given values of the covariate for the entire spatial region and estimates $\widehat{\boldsymbol{\alpha}}$, $\widehat{\boldsymbol{\delta}}$, $\widehat{\bf \Psi}$, and $\widehat{\bf \Gamma}$, correlation plots can be made for various reference points of interest (as in \citealp{spde13}) which can illuminate how the length and direction of dependence changes over the spatial region. In addition, the process variance can be plotted to visualize how variability in the spatial process changes with covariate information.

\section{A Bayesian nonstationary spatial model} \label{modelstatement}

A fully Bayesian model for a univariate spatial Gaussian process can now be defined as follows. Keeping the same notation as before, define $Y({\bf s})$ to be a mean-zero GP with the nonstationary parametric regression covariance function $C^R$ in (\ref{regression_covfcn}). However, now suppose we observe $Z({\bf s})$, a mean-adjusted noisy version of $Y({\bf s})$. Then, the model can be specified as 
\begin{equation} \label{model}
Z({\bf s}) = \mu({\bf s}) + Y({\bf s}) + \epsilon({\bf s}), 
\end{equation}
where $E[Z({\bf s})] = \mu({\bf s})$ is a deterministic mean function, the $\epsilon(\cdot)$ represents measurement error or microscale variability and is independent and identically distributed as $\mathcal{N}(0, \tau^2)$ with $\tau^2$ unknown, and $\epsilon(\cdot)$ and $Y(\cdot)$ are independent. (In general, $\mathcal{N}(a, b)$ is the univariate Gaussian distribution with mean $a$ and variance $b$.) It follows that for a fixed, finite set of $n$ spatial locations $\{{\bf s}_1, ... , {\bf s}_n\}\in G$, the random (observed) vector ${\bf Z} = \left( Z({\bf s}_1), ... , Z({\bf s}_n) \right)'$ will have a multivariate Gaussian distribution, and conditional on the other parameters in the model, the process ${\bf Y} = \left( Y({\bf s}_1), ... , Y({\bf s}_n) \right)'$ is distributed as $\mathcal{N}_n\big({\bf 0, \Omega}\big)$, where $\mathcal{N}_n\big({\bf a, B}\big)$ is the $n$-dimensional Gaussian distribution with mean vector ${\bf a}$ and covariance matrix ${\bf B}$. The elements of ${\bf \Omega}$ are $\Omega_{ij} \equiv C^R({\bf s}_i, {\bf s}_j; \boldsymbol{\alpha}, \boldsymbol{\delta}, {\bf \Psi, \Gamma})$.

\subsection{Computational details}

Model fitting for (\ref{model}) will be discussed for the usual spatial setting where $d=2$. Two additional simplifications will be made: first, the deterministic mean function will be a linear function of covariate information, i.e., $\mu({\bf s}) = {\bf x(s)}'\boldsymbol{\beta}$, where the mean coefficient vector has $dim(\boldsymbol{\beta}) = q$; second, the smoothness will be fixed to be non-spatially-varying and also constant, $\nu = 0.5$, so that the underlying correlation structure is exponential. A re-parameterization which will aide computation is to separate $\boldsymbol{\alpha} = (\alpha_0, \dots, \alpha_p)'$ into $\alpha_0$ and $\boldsymbol{\alpha}_{-0} = (\alpha_1, \dots, \alpha_p)'$, and then set $\sigma^2_0 = \exp\{ \alpha_0 \}$. The $\sigma^2_0$ term, a ``baseline variance,'' is present in every element of the covariance matrix ${\bf \Omega}$, and thus can be factored out and sampled in a Gibbs step. After these changes, the full parameter vector to be estimated is $\boldsymbol{\theta} = \big(\boldsymbol{\beta}, \tau^2, \sigma^2_0, \boldsymbol{\alpha}_{-0}, {\bf \Psi, \Gamma}\big)$.

\subsubsection{Prior specification}

We assume that the static model parameters $\boldsymbol{\theta}$ are \textit{a priori} independent, 
\begin{equation*}
p(\boldsymbol{\theta}) = p(\boldsymbol{\beta}) \> p(\tau^2) \> p(\sigma^2_0) \> p(\boldsymbol{\alpha}_{-0}) \> p({\bf \Psi}) \> p({\bf \Gamma}),
\end{equation*}
and specify proper prior distributions as follows. Since closed-form updates can be derived for the nugget, baseline variance, and mean coefficient parameters (see Appendix \ref{MCMC}), these will be given conjugate priors, namely $p(\tau^2) = \mathcal{IG}(a_\tau, b_\tau)$ (here, $\mathcal{IG}(a, b)$ is the inverse-Gamma distribution with shape $a$ and rate $b$), $p(\sigma_0^2) = \mathcal{IG}(a_\sigma, b_\sigma)$, and $p(\boldsymbol{\beta}) = \mathcal{N}_q({\bf 0}, c^2_\beta {\bf I}_q)$. Since closed-form full conditional distributions cannot be derived for the other parameters regardless of the form of the prior distribution, we specify $p(\boldsymbol{\alpha}_{-0}) = \mathcal{N}_{p-1}({\bf 0}, c^2_\alpha {\bf I}_{p-1})$ and $p({\bf \Gamma}) = p(\text{vec} \>{\bf \Gamma}) = \mathcal{N}_{dp}({\bf 0}, c^2_\Gamma {\bf I}_{dp})$. Finally, ${\bf \Psi}$ will be represented by its three unique parameters $(\psi_{11}, \psi_{22}, \rho)$, where $\rho = \psi_{12}/\sqrt{\psi_{11}\psi_{22}}$. This parameterization aides in specifying a noninformative prior for ${\bf \Psi}$: the diagonal elements $\psi_{11}$ and $\psi_{22}$ will be assigned diffuse half-Cauchy (positive-only) prior distributions, $p(\psi_{11}) = \mathcal{C}(0, s_1^2)$ and $p(\psi_{22}) = \mathcal{C}(0, s_2^2)$  (following \citealp{gelman2006}), where $\mathcal{C}(c, d)$ is the Cauchy distribution centered at $c$ with scale parameter $d$, and the correlation $\rho$ will be given a uniform prior over $(-1, 1)$. All hyperparameters are chosen so that the prior distributions are vague but proper.

\subsubsection{MCMC}

As usual, in this Bayesian setting, the posterior distribution cannot be obtained in closed form, and we  use to Markov chain Monte Carlo (MCMC) methods to obtain samples of the joint posterior. However, due to the conjugate prior specification, full conditional distributions for $\boldsymbol{\beta}, \tau^2$, and $\sigma^2_0$ can be obtained so that Gibbs steps can be used for these parameters. Componentwise random walk Metropolis-Hastings steps will be used for the covariance parameters $(\boldsymbol{\alpha}_{-0}, {\bf \Psi, \Gamma})$, although we note that multivariate Metropolis updates might be required as $p$ increases. For a full outline of the MCMC algorithm, see Appendix \ref{MCMC}.

\subsubsection{Posterior prediction}

Predictions at unobserved locations can be obtained as follows: define ${\bf Z}$ to be the $n$ observed process values and now define ${\bf Z^*}$ to be the values at $J$ unobserved locations. The posterior predictive distribution of interest is
\begin{equation} \label{ppd}
p ({\bf Z^* | z}) = \int_{\boldsymbol{\theta}} p({\bf Z^*, \boldsymbol{\theta} | z}) d\boldsymbol{\theta} =  \int_{\boldsymbol{\theta}} p({\bf Z^* | \boldsymbol{\theta}, z}) p(\boldsymbol{\theta}|{\bf z}) d\boldsymbol{\theta}. 
\end{equation}
Following the model specification in (\ref{model}),
\begin{equation*}
\left[ \begin{array}{c} {\bf Z} \\ {\bf Z^*} \end{array} \Bigg| \hskip1ex \boldsymbol{\theta} \hskip1ex \right] \sim \mathcal{N}_{n+J} \left( \left[ \begin{array}{c} {\bf {X}\boldsymbol{\beta}} \\ {\bf {X}^*\boldsymbol{\beta}} \end{array} \right] ,  \left[ \begin{array}{cc} \tau^2{\bf I}_n + {\bf \Omega_Z} & {\bf \Omega_{ZZ^*}} \\  {\bf \Omega_{Z^*Z}} & \tau^2{\bf I}_J + {\bf \Omega_{Z^*}}  \end{array} \right] \right),
\end{equation*}
so by conditional properties of the multivariate Gaussian distribution,
\begin{equation} \label{Zstar}
{\bf Z^* | Z = z}, \boldsymbol{\theta} \sim \mathcal{N}_J (\boldsymbol{\mu}_{\bf Z^*|z} , {\bf \Sigma}_{\bf Z^*|z} ),
\end{equation}
where
$\boldsymbol{\mu}_{\bf Z^*|z} =  {\bf {X}^*\boldsymbol{\beta}} - {\bf \Omega_{Z^*Z}} (\tau^2{\bf I}_n + {\bf \Omega_{Z}})^{-1} ({\bf z} - {\bf X}\boldsymbol{\beta})
$ and
${\bf \Sigma}_{\bf Z^*|z} = ( \tau^2{\bf I}_J + {\bf \Omega_{Z^*}}) - {\bf \Omega_{Z^*Z}} (\tau^2{\bf I}_n + {\bf \Omega_{Z}})^{-1}  {\bf \Omega_{ZZ^*}}. 
$ The integral in (\ref{ppd}) is not available in closed form, but given MCMC samples from the posterior $p(\boldsymbol{\theta}|{\bf z})$, say $\{ \boldsymbol{\theta}_l, l = 1, 2, ... , L \}$, we can compute a Monte Carlo estimate of the 
posterior predictive mean
\begin{equation} \label{ppm}
\widehat{E[{\bf Z^* | z}]} = L^{-1} \sum_{l=1}^L {\bf Z}^*_l,
\end{equation}
where the ${\bf Z}^*_l$ are draws from the distribution $[{\bf Z^* | z}, \boldsymbol{\theta}_l]$ from (\ref{Zstar}). Other inferential quantities, such as $(1-\alpha)100\%$ posterior predictive intervals, can be calculated by finding the $(100\alpha/2)^{\text{th}}$ and $(100(1-\alpha/2))^{\text{th}}$ percentiles of the $\{ {\bf Z}^*_l; l=1, ... , L \}$.

\section{Application: annual precipitation in Colorado, USA} \label{application}
\begin{figure}[!t] 
\begin{center}
\includegraphics[width=\textwidth]{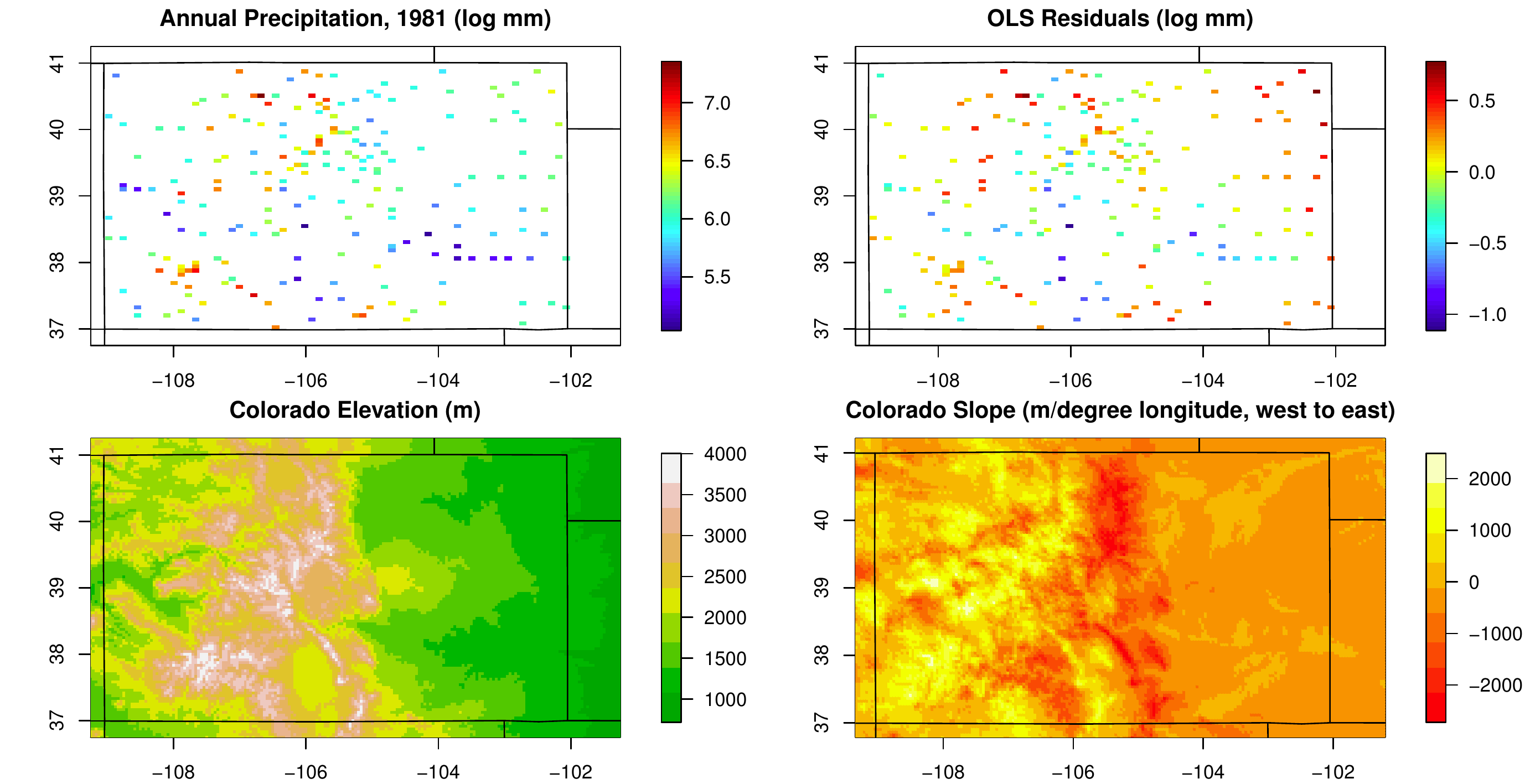}
\caption{Observation stations ($n=217$), labeled by log annual precipitation total (upper left) and residuals from a simple linear regression of log annual precipitation on elevation, slope, and the slope/elevation interaction (upper right); topographical map of Colorado with elevations in meters (lower left); a representation of the change in elevation or slope, measured as a west-to-east gradient (lower right).}
\label{data_elev}
\end{center}
\end{figure}

{\sloppypar
\subsection{Data} \label{data}

As an illustration of the regression-based nonstationary spatial model (\ref{model}), we analyze the precipitation dataset used by \cite{PacScher} from Colorado, a state in the western United States of America. Meteorological data from Colorado is available online from the National Center for Atmospheric Research at {\tt http://www.image.ucar.edu/GSP/Data/US.monthly.met/CO.html}. Specifically, the data used here consists of monthly precipitation recorded at each of approximately 400 weather stations in Colorado. We follow Paciorek and Schervish and consider annual precipitation recorded at the weather stations, choosing the 1981 records as they have the most stations (217) without missing monthly values. Precipitation has been recorded in millimeters, and the annual totals are transformed to be on the log scale in order to make the Gaussian process assumption more reasonable. The dataset used for analysis, which included precipitation, latitude, and longitude, was obtained upon request from Dr. Chris Paciorek.

Covariate information can be included in three parts of the nonstationary model: the mean function, the spatial variance function, and the kernel matrix function. Elevation is perhaps the most intuitive covariate to include in a covariance function for modeling precipitation in Colorado and provides good motivation for thinking about how covariate information might impact spatial dependence properties. A dramatic ridge of the Rocky Mountain range runs through the western part of Colorado, while the eastern part of Colorado is primarily flat. As a result, the topography is quite diverse: the entire eastern third of the state sits at about 1000 meters above sea level, while elevations in the western part of the state are highly variable and range from 1000 all the way up to 4000 meters above sea level. Intuitively, it is reasonable to think about elevation being a driving force for the nonstationarity in annual precipitation. Elevation measurements for each observation location were obtained from GPS Visualizer ({\tt http://www.gpsvisualizer.com/elevation}), an online tool, although elevation measurements for the entire state are available in the {\tt fields} package in {\tt R} (\citealp{R_fields}). In addition to elevation, another potentially useful covariate may be the change in elevation, or slope, at each location. The predominant weather patterns in Colorado move from west to east, and therefore it might be the case that the behavior of precipitation on west-facing mountainsides is quite different than the behavior on east-facing mountainsides. One way to measure slope that might be useful in this application is through a west-to-east gradient. The slope for each location is calculated by first obtaining the elevation of points at a fixed distance directly east and directly west of that location, and then subtracting the elevation to the west from the elevation to the east. We chose the fixed distance to be $5/6$ of a longitude unit, as this distance yielded a map that was neither overly smooth nor overly coarse. Slope measurements for the entire state were calculated using the state-wide elevation measurements provided in the {\tt fields} package, and the slope measurements for each observation location were picked out from the closest available grid cell from the full map.  See Figure \ref{data_elev} for a complete map of the elevation and slope.
}

Since the influence of elevation is clear in the raw data, a plot of the residuals from a standard ordinary least squares (OLS) procedure is also included in Figure \ref{data_elev}. The residuals are obtained by detrending the raw data with a mean structure that includes elevation, slope, and the interaction between elevation and slope.  In this plot, it is clear that there is still a strong degree of spatial structure in the data, even after correcting the mean with covariate information. Although we later take a fully Bayesian approach to estimating the mean and covariance parameters simultaneously, each of these covariates were found to be statistically significant in a classical preliminary analysis which made the standard regression assumptions of independence and error homogeneity, and therefore this mean model was chosen as a candidate for the spatial model. 
However, since it is much less straightforward to determine the significance of variance or kernel covariates \textit{a priori}, a ``full model'' will be fit to each of these components, such that both the variance and kernel structure will include elevation, slope, and their interaction. Elevation and slope measurements will be standardized in order to put the coefficient estimates on a similar scale. As discussed in Section \ref{sec:properties}, after fitting the full model, posterior credible intervals for the corresponding coefficients can be used to determine the importance of including each covariate in a final model. 

\subsection{Model comparison}

\subsubsection{Alternative models} \label{altmodels}

For evaluative purposes, the full nonstationary model introduced in Section \ref{data} will be compared to a simpler, second-order stationary Mat\'{e}rn (exponential) model. The stationary model used will simply be the nonstationary model (\ref{model}) with the variance coefficients $\boldsymbol{\alpha}_{-0}$ fixed to be zero, the smoothness again fixed at $\nu = 0.5$, and the kernel matrix for each location fixed to be constant (i.e., ${\bf \Sigma(s_i) \equiv \Sigma_0}$ for all ${\bf s_i}$). The anisotropy matrix ${\bf \Sigma_0}$ will be parameterized through a spectral decomposition by the vector $(\lambda_1, \lambda_2, \eta)$, where
\begin{equation} \label{eigen}
{\bf \Sigma_0} = \left[ \begin{array}{cc} \cos(\eta) & - \sin(\eta)  \\ \sin(\eta) & \cos(\eta)  \end{array} \right] \left[ \begin{array}{cc} \lambda_1 & 0  \\ 0 & \lambda_2  \end{array} \right] \left[ \begin{array}{cc} \cos(\eta) & \sin(\eta)  \\ -\sin(\eta) & \cos(\eta)  \end{array} \right].
\end{equation}
The parameters $\lambda_1$ and $\lambda_2$ can be interpreted as squared spatial ranges and will be assigned diffuse half-Cauchy prior distributions $\mathcal{C}(0, s_1^2)$ and $\mathcal{C}(0, s_2^2)$ (again because no closed-form updates can be derived regardless of the prior choice) with the same hyperparameters as those used for the diagonal elements of ${\bf \Psi}$ in the nonstationary model; $\eta$ can be interpreted as an angle of rotation for the anisotropy matrix and will be given a uniform prior over $[0, \pi/2]$ for identifiability purposes (following, e.g., \citealp{katzfuss2013}). 

As will be seen in Section \ref{results}, not every covariate will prove to be important in explaining nonstationarity in the spatial variance and kernel matrix processes. That is, many of the 95 percent credible intervals for the regression coefficients cover zero. Therefore, we consider a reduced model based on removing components in the full model with credible intervals that include zero. The resulting simplified model includes a constant variance and only elevation in the kernel matrix process. The mean function will remain unchanged. 

For notational simplicity, the stationary model will be labelled S-M1, the full nonstationary model will be labelled FNS-M2, and the reduced nonstationary model will be labelled RNS-M3.
 
\subsubsection{Evaluation criteria}

All three models given in Section \ref{altmodels} will be compared in terms of out-of-sample prediction. Ten percent of the full data ($J = 22$) will be held out from the 217 observations as test data (denoted ${\bf Z}_{\text{test}}$). The remaining $n=217-22=195$ observations will be used a training data (denoted ${\bf Z}_{\text{train}}$) to fit each of the models and predict at these $J$ locations. Using these predictions, three evaluation criteria will be used to compare the stationary and nonstationary models.

First, we will calculate mean squared prediction error
\begin{equation*}
MSPE = \frac{1 }{ J  }  \sum_{j=1}^J (z^*_j - \hat{z}^*_j)^2,
\end{equation*}
where $z^*_j$ is the $j$th held-out observed value and $\hat{z}^*_j$ is the corresponding predicted posterior mean (from (\ref{ppm})). Smaller MSPE indicates better predictions. 

Second, for a more formal comparison, the continuous rank probability score will be used (a proper scoring rule; see \citealp{PropScoring}). For the $j$th prediction, this is defined as
\begin{equation} \label{CRPS}
CRPS_j \equiv CRPS(F_j, z^*_j) = - \int_{-\infty}^{\infty} \big( F_j(x) - \mathbbm{1}  \{ x \geq z^*_j \} \big)^2dx,
\end{equation}
where $F_j(\cdot)$ is the cumulative distribution function for the predictive distribution of $z^*_j$ given ${\bf Z}_{\text{train}}$ and $\mathbbm{1}\{ \cdot \}$ is the indicator function. A Monte Carlo estimate of the CRPS can be obtained by averaging over the posterior samples obtained from the MCMC algorithm, 
\[
\widehat{CRPS}_j = - \frac{1}{L} \sum_{l=1}^L \int_{-\infty}^{\infty} \big( F_j(x; \boldsymbol{\theta}_l) - \mathbbm{1}  \{ x \geq z^*_j \} \big)^2dx,
\]
where $\{ \boldsymbol{\theta}_l, l = 1, 2, ... , L \}$ are the posterior samples and $F_j(\cdot; \boldsymbol{\theta}_l)$ is the conditional univariate (Gaussian) predictive cumulative distribution function given in (\ref{Zstar}) with $\boldsymbol{\theta} = \boldsymbol{\theta}_l$. In this case, given that the predictive CDF is conditionally Gaussian, a computational shortcut can be used for calculating (\ref{CRPS}): when $F$ is Gaussian with mean $\mu$ and variance $\sigma^2$,
\begin{equation*} \label{CRPS_normal}
CRPS \big(  F,  z^*_j \big) = - \sigma \Bigg[ \frac{1}{\sqrt{\pi}} - 2\cdot \phi\left(\frac{z^*_j - \mu}{\sigma} \right) - \frac{z^*_j - \mu}{\sigma} \left( 2 \cdot \Phi\left( \frac{z^*_j - \mu}{\sigma} \right) - 1 \right) \Bigg],
\end{equation*}
where $\phi$ and $\Phi$ denote the probability density and cumulative distribution functions, respectively, of a standard Gaussian random variable. The reported metric will be the average over all holdout locations,
$
\widehat{CRPS} = J^{-1}  \sum_{j=1}^J \widehat{CRPS}_j.
$
Larger CRPS indicates better model fit. 

Finally, we will use the logarithmic score, defined as
\begin{equation*}
\text{logScore} = \log \big( \hat{p} ({\bf Z^* | z}) \big)= \log \left( \frac{1}{L} \sum_{l=1}^L p({\bf Z^*} | \boldsymbol{\theta}_l, {\bf z}) \right)
\end{equation*}
(\citealp{good1952}). A larger logarithmic score indicates better model fit.

Instead of evaluating the criteria for a single hold-out sample, 20 sets of hold-out samples will be taken and the evaluation criteria will be calculated for each to account for the variability in prediction accuracy implicit in the choice of a particular hold-out sample. Lastly, note that information criterion based methods (such as deviance information criterion) should not be used as a model selection tool since we are comparing models that have different ranks (\citealp{hoff_niu}).  The stationary model is a full-rank representation of the kernel matrix process, while the nonstationary model is a rank-1 representation (see Section \ref{discussion} for further discussion).

\subsection{Results} \label{results}

As a consequence of the low-dimensional parameter space, the MCMC algorithm appears to converge quickly, within 500 iterations. Nevertheless, the algorithm was run for 10,000 iterations in order to ensure convergence, which was assessed through examination of marginal trace plots for each model parameter. The hyperparameters were fixed to $c_\beta^2 = 100^2$, $a_\tau = 2$, $b_\tau = 0.05$, $a_\sigma = 2$, $b_\sigma = 0.5$, $c_\alpha^2 = 100$, $s_1^2 = s_2^2 = 1$, $c_\Gamma^2 = 5$.

Posterior summaries of parameters in models S-M1 and FNS-M2 are summarized in Table \ref{regressiontable}. The posterior means for the mean coefficients ($\boldsymbol{\beta}$) are nearly equal in these two models, as is the point estimate of the nugget variance ($\tau^2$). In the process variance and kernel matrix functions for FNS-M2, the coefficients corresponding to the interaction covariate ($\alpha_3$, $\gamma_{14}$, and $\gamma_{24}$) have credible intervals that solidly include zero, which indicates that the elevation/slope interaction is not useful in explaining changes over space in the process variance and kernel matrices. As a result, we only interpret the marginal effects of elevation and slope for the process variance and kernel matrix functions. 

In the variance function, the posterior mean of the elevation coefficient ($\alpha_1$) is positive, indicating that higher elevations correspond to larger variance; the variance increases by an estimated multiplicative factor of $\exp \{ 0.147 \} = 1.158$ for each additional unit of (standardized) elevation. The posterior mean for the slope coefficient ($\alpha_2$) is also positive, estimating that the variance increases by a multiplicative factor of $\exp \{ 0.101 \} = 1.106$ for each additional unit of (standardized) slope. Recall the prevailing weather patterns discussed in Section \ref{data}: this indicates that the steep west-facing mountainsides which experience the full force of a weather system are estimated to have larger variability in precipitation than the steep east-facing mountainsides which are sheltered from storms approaching from the west. However, the 95\% credible interval for both the elevation and slope coefficients includes zero, so these covariates are likely not important in explaining variability in the process variance. It is interesting to note that the posterior mean estimate for the baseline variance parameter ($\sigma^2_0$) is much smaller in the nonstationary model, so while the covariates may not be important for explaining variability, the model is still using them to allow the variance to change over space. 

In the kernel matrix function, the elevation covariate is important for explaining variability in the kernels matrices, albeit only in the $x$ direction -- the credible interval for $\gamma_{12}$ excludes zero, while the coefficient for the $y$ direction ($\gamma_{22}$) has a credible interval that includes zero (barely). Holding all other covariates constant, based on the posterior mean for these parameters, the minimum kernel variance in the $x$-direction as a function of elevation occurs when the standardized elevation covariate is $-\hat{\gamma}_{11}/\hat{\gamma}_{12} = 0.292/1.77 = 0.165$ (following the interpretations given in Section \ref{Kern_Reg}). Similarly, the minimum kernel variance in the $y$-direction as a function of elevation occurs when the standardized elevation covariate is $-\hat{\gamma}_{21}/\hat{\gamma}_{22} = -0.312/0.869 = -0.359$. The standardized elevation covariate ranges between about $-2$ and $2$, so these minima occur near the middle of this range. It follows that more extreme elevation values (both low and high) are estimated to have larger kernel variances in both directions. Similar interpretations could be made for the coefficients corresponding to slope, but both of these coefficients have credible intervals that include zero so again the slope covariate might not be important for explaining variability in the kernel matrix process.

\begin{figure}[!t] 
\begin{center}
\includegraphics[width=\textwidth]{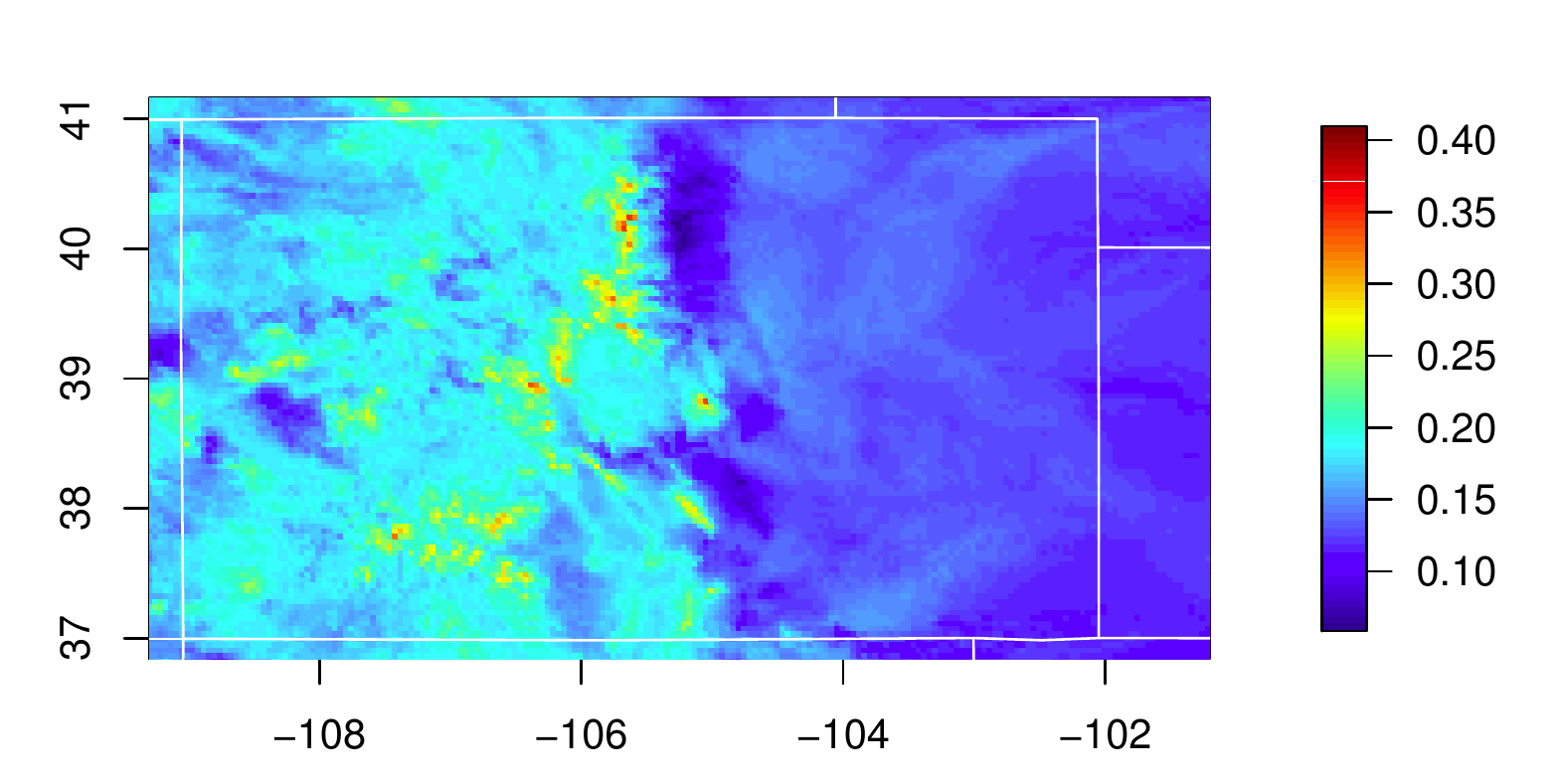}
\caption{A plot of the estimated spatially-varying process variance, calculated using the posterior mean of the parameters.}
\label{variance}
\end{center}
\end{figure}

\begin{figure}[!t] 
\begin{center}
\includegraphics[width=\textwidth]{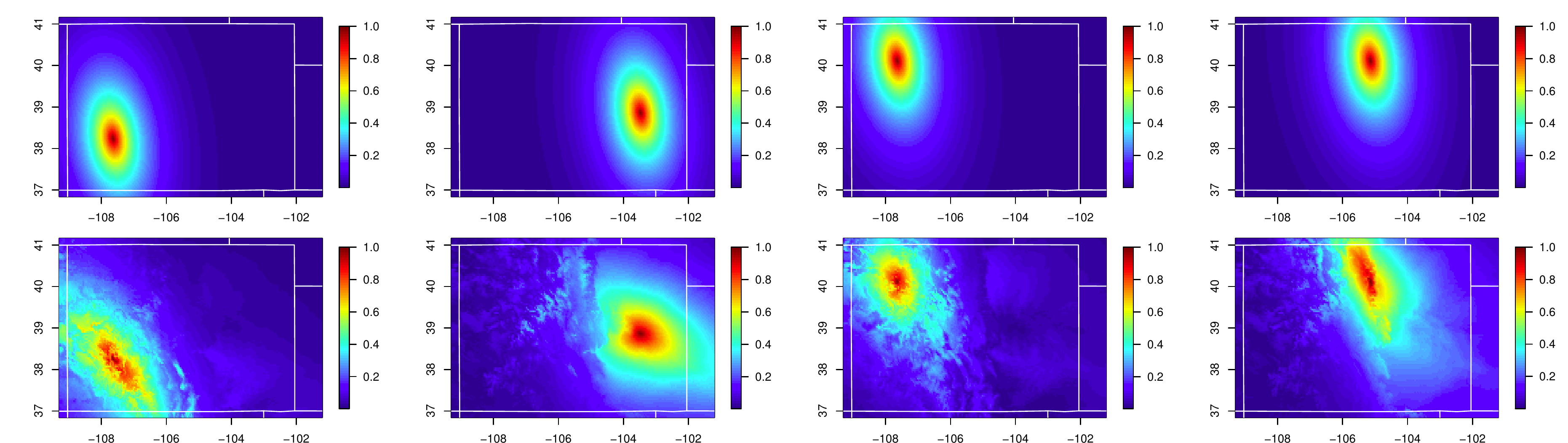}
\caption{Correlation plots for four reference points in the stationary model S-M1 (top) and nonstationary model FNS-M2 (bottom), calculated using the posterior mean parameter estimates.}
\label{corrplots}
\end{center}
\end{figure}

In addition to describing the extent to which covariates impact spatial dependence, another benefit of using the nonstationary regression model is that we can produce visual summaries of how the nonstationary behavior of the spatial process changes over space. First, using posterior mean estimates from the FNS-M2 model, consider a plot of how the point estimate of the process variance changes over space, given in Figure \ref{variance}. As expected (and as can be interpreted from the variance coefficients), the largest variance point estimate is in the high mountainous areas, but interestingly the smallest variance point estimate falls on the east-facing slopes of the mountains, not in the flat plains area. Perhaps a more interesting summary is obtained in the correlation plots in Figure \ref{corrplots}, which illustrate the spatially-varying anisotropy for four reference points over the spatial region. These plots are created using the posterior means for ${\bf \Psi}$ and ${\bf \Gamma}$. Presented alongside the correlation plots for the nonstationary model are the corresponding correlations in the stationary model S-M1, again calculated using the posterior mean of ${\bf \Psi}$. Clearly, this model can give correlation patterns which are non-elliptical and non-monotone.

\begin{figure}[!t] 
\begin{center}
\includegraphics[width=\textwidth]{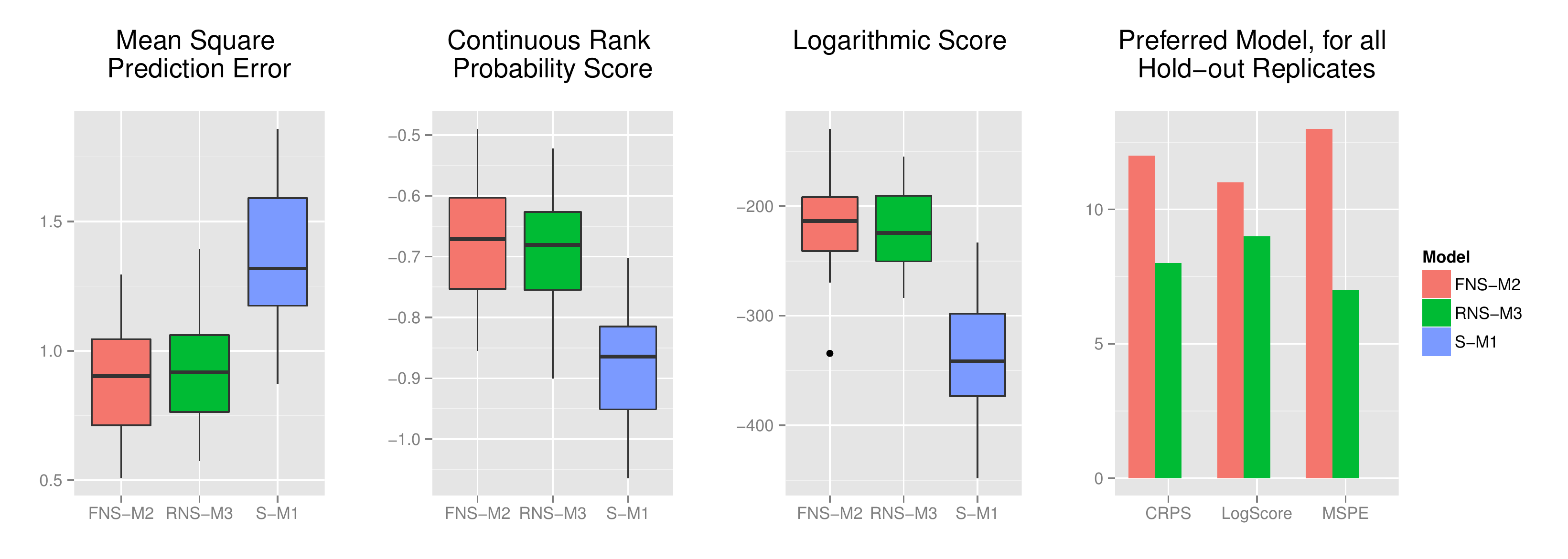}
\caption{Boxplots of the evaluation criteria for each of the models fit to the Colorado precipitation data, summarized for each of 20 holdout replicates. Recall: S-M1 is the stationary model, FNS-M2 is the full nonstationary model, RNS-M3 is the reduced nonstationary model. Small MSPE indicates better model fit; larger CRPS and log score indicate better model fit. The bar plot on the right summarizes which model is chosen as ``best'' under each criteria when the three models are compared separately for each of the 20 hold-out sets.}
\label{eval_crit}
\end{center}
\end{figure}

Finally, consider the distribution of the three of the evaluation criteria across the 20 sets of hold-out samples, which are summarized by boxplots in Figure \ref{eval_crit}. In this plot we compare each of the three models, S-M1, FNS-M2, and RNS-M3. In general, the full and reduced nonstationary models perform much better than the stationary model under all three criteria. Also in Figure \ref{eval_crit}, note that when the models are compared for each hold-out set separately, the full nonstationary model is most often preferred, and the stationary model is never preferred.

\section{Discussion} \label{discussion}

In this paper, we have presented a new class of regression-based nonstationary Mat\'{e}rn covariance functions, which (like mean regression functions) rely on observable covariate processes to explain the spatially-varying second-order properties of the spatial process of interest. As a result, our model is an extremely parsimonious representation of the spatial process and provides a practical compromise between stationary models and similar nonstationary models which are difficult to fit. The model is highly flexible and allows for potentially different covariates to model each of the mean, variance, smoothness, and anisotropy. Stationary models are a special case of our general nonstationary specification; furthermore, the model can indicate stationarity in the process even when it is not desired to assume as much. Finally, given parameter estimates, our model allows for interpretations and visualizations of the nonstationarity in the process, providing an explanation of why the process is nonstationarity.


Of course, the \cite{hoff_niu} model (\ref{kern_reg}) is not the only way to parameterize a kernel matrix so that it varies spatially according to covariate information. In the two-dimensional spatial case, a $2\times2$ kernel (covariance) matrix has three unique parameters, which can be related to covariates in a variety of ways. For example, following \cite{katzfuss2013}, the spectral decomposition of ${\bf \Sigma} = (\lambda_1, \lambda_2, \eta)$ could be used, as in (\ref{eigen}) (although even this choice is not unique for a spectral decomposition; see \citealp{PacScher}, or \citealp{Higdon99}). Alternatively, the kernel matrix could be parameterized more directly as ${\bf \Sigma} = (\sigma^2_1, \sigma^2_2, \rho)$, where $\rho = \sigma_{12}/(\sigma_1 \sigma_2)$. For either of these choices (or any other choice), an intuitive way to introduce a regression framework is to define generalized linear models for each of these three components which allow each parameter to vary spatially, and estimate the corresponding coefficients. A nonparametric approach to estimating the kernel process could be implemented, and indeed this is essentially the approach taken by \cite{PacScher}. However, as already discussed, introducing a nonparametric model for the kernel process is not at all a parsimonious representation, which leads to model fitting difficulties. In a similar approach, \cite{katzfuss2013} introduced a reduced-rank nonparametric basis function model for the spatial dependence parameters, which avoids the model fitting problems of \cite{PacScher}. Regardless, since both of these are stochastic models, it is not possible to recover a stationary model as a special case, and interpretability is difficult. Future work includes exploring a hybrid approach, which models spatially-varying parameters through both parametric (covariate-driven) and nonparametric components.

The primary reason for choosing the Hoff model is that the parameters represent the kernel matrices directly, as opposed to representing the kernel matrices on a transformed scale. Another reason is that Hoff's model automatically includes a quadratic term for each covariate; furthermore, it provides a more parsimonious representation when the covariate dimension $p$ increases. For $d=2$, the dimension of the parameter space for (\ref{kern_reg}) is $3 + 2p$; any of the other parameterizations mentioned above have a parameter space with dimension $3p$, which is great than or equal to $3+2p$ when $p\geq3$. A final reason for choosing the Hoff parameterization is also computational: the parameters that correspond to the covariate (${\bf \Gamma}$) have no restrictions, as they need only be real numbers. The difficult parameter to estimate in this model is the baseline kernel matrix ${\bf \Psi}$, and this matrix is sampled separately from the parameters associated with ${\bf x(s)}$ in the MCMC algorithm. In our trial implementations of the other two parametric representations suggested above, difficulties were encountered due to the fact that each model contains a parameter which must vary spatially but is defined on a bounded interval (the angle of rotation $\eta$ and the correlation $\rho$). The appropriate link function for these parameters is a scaled and shifted logit or probit function, and because the parameter might need to be exactly equal or very close to both endpoints of its interval on the raw scale ($\eta$ may need to be exactly 0 or $\pi/2$, and $\rho$ may need to be close to $-1$ or $1$) the coefficients on the transformed scale were unstable and difficult to estimate.

However, one limitation of the \cite{hoff_niu} representation is that the difference between the kernel matrix ${\bf \Sigma(s)}$ and the baseline kernel matrix ${\bf \Psi}$ is required to be a rank one matrix. That is, rearranging (\ref{kern_reg}) gives ${\bf \Sigma(s) - \Psi = \Gamma x(s) x(s)' \Gamma'}$, and $\text{rank} \big({\bf \Gamma x(s) x(s)' \Gamma'} \big) = 1$, an issue which is noted in a recent extension of the original paper (\citealp{Hoff_unpub}). This limitation was discussed in Section \ref{Kern_Reg}, in that the off-diagonal elements of ${\bf\Sigma(s)}$ are modeled with the same parameters as the diagonal elements. As mentioned in \cite{Hoff_unpub}, the model can be extended to give kernel matrices whose difference from the baseline is any desired rank. In the two-dimensional spatial case, the model can be brought up to full rank by adding an additional quadratic component, for example
$\tilde{\bf \Sigma}{\bf (s) = \Psi + \Gamma}_1{\bf x(s) x(s)' \Gamma}'_1 + {\bf \Gamma}_2{\bf x(s) x(s)' \Gamma}'_2$,
although this of course increases the dimension of the parameter space and hence increases the complexity of the model. 

While the model outlined in this paper is currently for spatial data sets only, it is straightforward to apply this modeling framework to spatio-temporal problems. The easiest case for which this can be done is when the covariate itself is temporally-varying (e.g., wind, as in \citealp{calder08}), since the covariate would control the change in spatial covariance over time. Another way to incorporate time without any changes to the parameter space of the model would be to include time as a covariate, which would ensure temporal smoothness in the kernel matrices in the same way that the kernel matrix process was found to be smooth for any other smooth covariate. Alternatively, the kernel parameters could themselves be temporally varying, e.g.,
$
{\bf \Sigma}_t({\bf s}) = {\bf \Psi}_t + {\bf \Gamma}_t {\bf x(s) x(s)' \Gamma}'_t$.

{\sloppypar
Regarding computational feasibility for this model, as discussed in Section \ref{results}, the low dimensional nature of the parameter space allows the MCMC algorithm to converge very quickly. As a result, the main limitations on scalability for this model lie in two areas: evaluating the Gaussian likelihood function and calculating the elements of the covariance matrix for particular parameter values. The training data set used in the application had approximately 200 observations, and running 10000 iterations of the MCMC algorithm on a 2x Eight Core Xeon E5-2680 machine with 2.7 GHz of memory and 384 GB of storage took approximately 20 hours. It should be noted that the MCMC algorithm could be made much more efficient: currently, univariate Metropolis-Hastings steps were used for each of the kernel parameters, meaning that for the full nonstationary model there were twelve likelihood calculations for each iteration. An algorithm that uses adaptive multivariate Metropolis-Hastings steps may greatly reduce the computational demands of fitting this model. Regardless of the model-fitting algorithm, more efficient methods for calculating the covariance matrix are being explored; parallel computing strategies could also be employed to expedite the calculations. Our future work explores these extensions.}

\clearpage

\section*{Tables}

\begin{table}[h]
\begin{center}
\begin{tabular}{c|l||cc||cc|}
 \cline{2-6}  						
 & \multirow{2}{*}{ \textbf{Coefficient}} & \multicolumn{2}{ |c|| }{ { \textbf{S-M1} } }& \multicolumn{2}{ |c| }{ { \textbf{FNS-M2 } } }  \\  \cline{3-6}
  & & Mean & 95\% CI  & Mean &  95\% CI   \\
\cline{2-6} 
\hline
\multicolumn{1}{ |c| }{ \multirow{4}{*}{ \rotatebox[origin=c]{90}{ \textit{Mean} } } }& $\beta_0$ (intercept) & \textbf{6.486}  & \textbf{(6.349, 6.692)}  &  \textbf{6.308}  &  \textbf{ (6.155, 6.488) } \\ \cline{2-6}
\multicolumn{1}{ |c| }{} & $\beta_1$ (elevation) & \textbf{0.546}  & \textbf{(0.511, 0.579)}  &  \textbf{ 0.477 }  &  \textbf{(0.384, 0.575)}  \\ \cline{2-6} 
\multicolumn{1}{ |c| }{} & $\beta_2$ (slope) & \textbf{0.033}  & \textbf{(0.018, 0.047)}  & \textbf{0.053}  & \textbf{(0.013,  0.091)} \\ \cline{2-6} 
\multicolumn{1}{ |c| }{} & $\beta_3$ (interaction) & \textbf{0.054}  & \textbf{(0.044, 0.065)} &  \textbf{0.074} &  \textbf{(0.022,  0.124)}\\  
\hline
\hline
\multicolumn{1}{ |c| }{ \multirow{4}{*}{ \rotatebox[origin=c]{90}{ \textit{Variance} } } } & $\sigma^2_0 = \exp\{ \alpha_0 \}$ & 0.202  & (0.121, 0.342)  & 0.163 & (0.115, 0.227)  \\ \cline{2-6}
\multicolumn{1}{ |c| }{} & $ \alpha_1 $ (elev.) & --  & --  & 0.147 & (-0.127,  0.410)  \\ \cline{2-6} 
\multicolumn{1}{ |c| }{} &  $ \alpha_2 $ (slope) & --  & --  & 0.101  &  (-0.089, 0.308)  \\ \cline{2-6} 
\multicolumn{1}{ |c| }{} & $ \alpha_3 $ (int.) & --  & --  & -0.124 &  (-0.391, 0.126) \\  
\hline
\hline
\multicolumn{1}{ |c| }{ \multirow{8}{*}{ \rotatebox[origin=c]{90}{ \textit{Kernel} } } } & $\gamma_{11}$ (intcpt. 1) & --  & --  										& -0.292 	& (-1.429, 0.756) \\ \cline{2-6}
\multicolumn{1}{ |c| }{} & $\gamma_{12}$ (elev. 1) & --  & --   &\textbf{1.770}  & \textbf{(0.420, 3.310)} \\ \cline{2-6} 
\multicolumn{1}{ |c| }{} & $\gamma_{13}$ (slope 1)  & -- & --  & -0.752    & (-1.723, 0.200)  \\ \cline{2-6} 
\multicolumn{1}{ |c| }{} & $\gamma_{14}$ (int. 1) & --  & -- 	    &0.571 	& (-0.394, 1.543)  \\  \cline{2-6}
\multicolumn{1}{ |c| }{} & $\gamma_{21}$ (intcpt. 2) & --  & --  &-0.312 	&(-1.278, 0.596) \\ \cline{2-6} 
\multicolumn{1}{ |c| }{} & $\gamma_{22}$ (elev. 2) & --  & --    &{-0.869} & {(-1.981, 0.063)} \\ \cline{2-6} 
\multicolumn{1}{ |c| }{} & $\gamma_{23}$ (slope 2)  & --  & --  	&0.876 	&(-0.687, 2.034)  \\ \cline{2-6} 
\multicolumn{1}{ |c| }{} & $\gamma_{24}$ (int. 2) & --  & -- 	&0.134 	& (-1.053, 1.952)  \\  
\hline
\hline
\multicolumn{1}{ |c| }{ \multirow{4}{*}{ \rotatebox[origin=c]{90}{ \textit{Other} } } } & $\psi_{11}$ & 2.314  & (0.624, 6.633)   & 0.602 & (0.262, 1.215) \\ \cline{2-6}
\multicolumn{1}{ |c| }{} & $\psi_{22}$  & 3.430 & (0.927, 9.330)  &  1.240 & (0.535, 2.671)  \\ \cline{2-6} 
\multicolumn{1}{ |c| }{} & $\psi_{12}$  & -0.748  & (-2.965, 0.241)  & -0.153 & (-0.737, 0.354)  \\ \cline{2-6} 
\multicolumn{1}{ |c| }{} & $\tau^2$ (nugget) & 0.012  & (0.006, 0.020) & 0.010 & (0.006, 0.016) \\  
\hline
\end{tabular}
\end{center}
\caption{Posterior means and 95\% credible intervals for the stationary (S-M1) and full nonstationary (FNS-M2) models. Coefficient estimates in \textbf{bold} indicate those with a 95\% credible interval that does not include zero. Note: all covariates have been standardized.}
\label{regressiontable}
\end{table}

\clearpage

\section*{Acknowledgements}

We acknowledge helpful discussions with Veronica Berrocal and Michael Stein in the development of this research. This work was supported in part by the Statistical Methods for Atmospheric and Oceanic Sciences (STATMOS) research network (NSF-DMS awards 1106862, 1106974, and 1107046).

\section*{Supplementary materials}

Additional supporting information, including the Matlab code used to fit this model and the 1981 annual  precipitation data set from Colorado (with elevation, slope, and holdout indicator variables), may be found online at the publisher's website.

\bibliographystyle{apalike} \bibliography{MDRResearch}

\appendix

\section{Markov chain Monte Carlo algorithm} \label{MCMC}

A full specification of the algorithm used to fit the general model (\ref{model}) is as follows. In a Bayesian framework such as this, we are able to obtain a number of conditional distributions in closed form, and can therefore use what \cite{vanDyk1} call partially collapsed Gibbs sampling methods. Using the generic notation that $[U| V]$ represents the conditional distribution of $U$ given $V$, these closed form distributions are as follows:
\begin{equation} \label{sample_beta}
[\boldsymbol{\beta} | {\bf Z}, \tau^2, \sigma^2_0, \boldsymbol{\alpha}_{-0}, {\bf \Psi, \Gamma}] = 
\mathcal{N}_q\left( {\bf \Sigma_{\boldsymbol{\beta}}} {\bf {X}}'(\tau^2{\bf I}_n + {\bf \Omega})^{-1} {\bf z}, {\bf \Sigma_{\boldsymbol{\beta}}}  \right)
\end{equation}
where ${\bf \Sigma_{\boldsymbol{\beta}}} = \Big[c_{\beta}^{-2}{\bf I}_q + {\bf {X}}'(\tau^2{\bf I}_n + {\bf \Omega})^{-1} {\bf {X}}  \Big]^{-1} = c_{\beta}^{2}{\bf I}_q - c_{\beta}^{4} {\bf {X}}'\Big[\tau^2{\bf I}_n + {\bf \Omega} + c_\beta^2 {\bf {X}}{\bf {X}}' \Big]^{-1} {\bf {X}} $ (using the Sherman-Morrison-Woodbury identity),
\begin{equation} \label{sample_tausq}
[\tau^2 | {\bf Y, Z}] = \mathcal{IG}\left( a_{\tau} + \frac{n}{2} , b_{\tau} + \frac{1}{2}\sum_{i=1}^n(z_i - y_i)^2 \right)
\end{equation}
\begin{equation} \label{sample_sigmasq}
[\sigma^2_0 |  {\bf Y}, \boldsymbol{\beta}, \boldsymbol{\alpha}_{-0}, {\bf \Psi, \Gamma}] = \mathcal{IG}\left( a_{\sigma} + \frac{n}{2} , b_{\sigma} + \frac{1}{2}({\bf Y} - {\bf {X} \boldsymbol{\beta}})'{\bf \tilde{\Omega}}^{-1}({\bf Y} - {\bf {X} \boldsymbol{\beta}}) \right)
\end{equation}
Here, ${\bf \tilde{\Omega}} \equiv {\bf \tilde{\Omega}}(\boldsymbol{\alpha}_{-0}, {\bf \Psi, \Gamma)}$ is the unscaled covariance matrix (i.e., not including the $\sigma^2_0$ term). Finally,
\begin{equation} \label{sample_Y}
[{\bf Y  | Z }, \boldsymbol{\theta}] = \mathcal{N}_n\Big({\bf \Sigma_Y} ({\bf \Omega}^{-1}{\bf {X} \boldsymbol{\beta}} + \tau^{-2} {\bf z}),  {\bf \Sigma_Y} \Big),
\end{equation}
where ${\bf \Sigma_Y} = ({\bf \Omega}^{-1} + \tau^{-2}{\bf I_n})^{-1} = {\bf \Omega} - {\bf \Omega}({\bf \Omega} + \tau^{2}{\bf I_n})^{-1}{\bf \Omega}$ is re-written using the Sherman-Morrison-Woodbury identity. 

The parent (proper) Gibbs sampler, which is not unique, is then as follows (\textbf{Sampler 1}):

\noindent\hrulefill

\noindent {\rm {\bf Sampler 1} }

\begin{enumerate}
\item Draw ${\bf Y}$ from $[{\bf Y |  Z}, \boldsymbol{\beta}, \tau^2, \sigma^2_0, \boldsymbol{\alpha}_{-0}, {\bf \Psi, \Gamma}]$
\item Draw $\boldsymbol{\beta}$ from $[\boldsymbol{\beta} | {\bf Y, Z}, \tau^2, \sigma^2_0, \boldsymbol{\alpha}_{-0}, {\bf \Psi, \Gamma}]$
\item Draw $\tau^2$ from $[\tau^2 | {\bf Y, Z}, \boldsymbol{\beta}, \sigma^2_0, \boldsymbol{\alpha}_{-0}, {\bf \Psi, \Gamma}] = [\tau^2 | {\bf Y, Z}]$
\item Draw $\sigma^2_0$ from $[\sigma^2_0 | {\bf Y, Z}, \boldsymbol{\beta}, \tau^2, \boldsymbol{\alpha}_{-0}, {\bf \Psi, \Gamma}] = [\sigma^2_0 | {\bf Y}, \boldsymbol{\beta}, \boldsymbol{\alpha}_{-0}, {\bf \Psi, \Gamma}]$
\item Draw $(\boldsymbol{\alpha}_{-0}, {\bf \Psi, \Gamma})$ from $[\boldsymbol{\alpha}_{-0}, {\bf \Psi, \Gamma} | {\bf Y, Z}, \boldsymbol{\beta}, \tau^2, \sigma^2_0]$
\end{enumerate}

\vskip-2ex

\noindent\hrulefill

First, we can marginalize, in which we move quantities in some steps from being conditioned upon to being sampled. Marginalization does not affect the stationary distribution of the chain, but it can sometimes provide computational gains in terms of convergence (\citealp{vanDyk1}). Following \cite{vanDyk1}, the superscript $\star$ will be used to denote a quantity that is resampled in a future step and is not part of the output of one iteration of the sampler. The quantity conditioned upon in any particular step will be the most recently sampled value of that quantity. This leads us to \textbf{Sampler 2}:

\noindent\hrulefill

\noindent {\rm {\bf Sampler 2} }

\begin{enumerate}
\item Draw ${\bf Y^{\star}}$ from $[{\bf Y |  Z}, \boldsymbol{\beta}, \tau^2, \sigma^2_0, \boldsymbol{\alpha}_{-0}, {\bf \Psi, \Gamma}]$
\item (\textbf{Marginalize}) Draw $\boldsymbol{\beta}, {\bf Y^{\star}}$ from $[\boldsymbol{\beta}, {\bf Y | Z}, \tau^2, \sigma^2_0, \boldsymbol{\alpha}_{-0}, {\bf \Psi, \Gamma}]$
\item Draw $\tau^2$ from $[\tau^2 | {\bf Y, Z}, \boldsymbol{\beta}, \sigma^2_0, \boldsymbol{\alpha}_{-0}, {\bf \Psi, \Gamma}] = [\tau^2 | {\bf Y, Z}]$
\item Draw $\sigma^2_0$ from $[\sigma^2_0 | {\bf Y, Z}, \boldsymbol{\beta}, \tau^2, \boldsymbol{\alpha}_{-0}, {\bf \Psi, \Gamma}] = [\sigma^2_0 | {\bf Y}, \boldsymbol{\beta}, \boldsymbol{\alpha}_{-0}, {\bf \Psi, \Gamma}]$
\item (\textbf{Marginalize}) Draw $(\boldsymbol{\alpha}_{-0}, {\bf \Psi, \Gamma}), {\bf Y}$ from $[\boldsymbol{\alpha}_{-0}, {\bf \Psi, \Gamma}, {\bf Y | Z}, \boldsymbol{\beta}, \tau^2, \sigma^2_0]$
\end{enumerate}

\vskip-2ex

\noindent\hrulefill

\vskip1ex

\noindent To set up the use of the third and final tool of the PCG sampler, we next permute the steps, which again does not alter the stationary distribution of the chain. This leads to \textbf{Sampler 3}:

\noindent\hrulefill

\noindent {\rm {\bf Sampler 3} }

\begin{enumerate}
\item (Formerly Step 2) Draw $\boldsymbol{\beta}, {\bf Y^{\star}}$ from $[\boldsymbol{\beta}, {\bf Y | Z}, \tau^2, \sigma^2_0, \boldsymbol{\alpha}_{-0}, {\bf \Psi, \Gamma}]$
\item (Formerly Step 5) Draw $(\boldsymbol{\alpha}_{-0}, {\bf \Psi, \Gamma}), {\bf Y^{\star}}$ from $[\boldsymbol{\alpha}_{-0}, {\bf \Psi, \Gamma}, {\bf Y | Z}, \boldsymbol{\beta}, \tau^2, \sigma^2_0]$
\item (Formerly Step 1) Draw ${\bf Y}$ from $[{\bf Y |  Z}, \boldsymbol{\beta}, \tau^2, \sigma^2_0, \boldsymbol{\alpha}_{-0}, {\bf \Psi, \Gamma}]$
\item (Formerly Step 3) Draw $\tau^2$ from $[\tau^2 | {\bf Y, Z}, \boldsymbol{\beta}, \sigma^2_0, \boldsymbol{\alpha}_{-0}, {\bf \Psi, \Gamma}] = [\tau^2 | {\bf Y, Z}]$
\item (Formerly Step 4) Draw $\sigma^2_0$ from $[\sigma^2_0 | {\bf Y, Z}, \boldsymbol{\beta}, \tau^2, \boldsymbol{\alpha}_{-0}, {\bf \Psi, \Gamma}] = [\sigma^2_0 | {\bf Y}, \boldsymbol{\beta}, \boldsymbol{\alpha}_{-0}, {\bf \Psi, \Gamma}]$
\end{enumerate}

\vskip-2ex

\noindent\hrulefill

\vskip1ex

\noindent However, this sampler is inefficient, as it samples ${\bf Y}$ three separate times. Since we have arranged the steps such that the intermediate quantities ${\bf Y^{\star}}$ are not used in subsequent steps (the only ${\bf Y}$ we condition on in the chain is the one drawn in step 3 of Sampler 3), we can remove or trim the intermediate draws from the sampler. This does not affect the transition kernel and will therefore leave the stationary distribution unaltered (\citealp{vanDyk1}). Here is the final sampler, denoted \textbf{Sampler 4}:

\noindent\hrulefill

\noindent {\rm {\bf Sampler 4 (Final sampler)} }

\begin{enumerate}
\item Draw $\boldsymbol{\beta}$ from $[\boldsymbol{\beta} | {\bf Z}, \tau^2, \sigma^2_0, \boldsymbol{\alpha}_{-0}, {\bf \Psi, \Gamma}]$
\item Draw $(\boldsymbol{\alpha}_{-0}, {\bf \Psi, \Gamma})$ from $[\boldsymbol{\alpha}_{-0}, {\bf \Psi, \Gamma} | {\bf Z}, \boldsymbol{\beta}, \tau^2, \sigma^2_0]$
\item Draw ${\bf Y}$ from $[{\bf Y |  Z}, \boldsymbol{\beta}, \tau^2, \sigma^2_0, \boldsymbol{\alpha}_{-0}, {\bf \Psi, \Gamma}]$
\item Draw $\tau^2$ from $[\tau^2 | {\bf Y, Z}, \boldsymbol{\beta}, \sigma^2_0, \boldsymbol{\alpha}_{-0}, {\bf \Psi, \Gamma}] = [\tau^2 | {\bf Y, Z}]$
\item Draw $\sigma^2_0$ from $[\sigma^2_0 | {\bf Y, Z}, \boldsymbol{\beta}, \tau^2, \boldsymbol{\alpha}_{-0}, {\bf \Psi, \Gamma}] = [\sigma^2_0 | {\bf Y}, \boldsymbol{\beta}, \boldsymbol{\alpha}_{-0}, {\bf \Psi, \Gamma}]$
\end{enumerate}

\vskip-2ex

\noindent\hrulefill

\vskip1ex

\noindent This sampler is preferred because a closed form for the distribution is available in steps 1, 3, 4, and 5 (equations (\ref{sample_beta}), (\ref{sample_Y}), (\ref{sample_tausq}), and (\ref{sample_sigmasq}) respectively) but still maintains the desired target stationary distribution. 

\subsubsection{Metropolis-Hastings steps for the kernel parameters}

Due to the form of the covariance function $C^R$, the closed-form full conditional distribution for $(\boldsymbol{\alpha}_{-0}, {\bf \Psi, \Gamma})$ is unavailable. Therefore, we use Metropolis-Hastings steps for step 2 of Sampler 4. The distribution we want to sample from in this step is $[\boldsymbol{\alpha}_{-0}, {\bf \Psi, \Gamma} | {\bf Z}, \boldsymbol{\beta}, \tau^2, \sigma^2_0]$, which uses the marginalized likelihood of $[{\bf Z} | \boldsymbol{\theta}] = \int [{\bf Z | Y}, \boldsymbol{\theta}][{\bf Y} | \boldsymbol{\theta}] d{\bf Y}$. 

Since the dimension of $(\boldsymbol{\alpha}_{-0}, {\bf \Gamma})$ is not too large (for the full model fit in this paper, the dimension is $3+8 = 11$), component-wise steps will be constructed for $\boldsymbol{\alpha}_{-0}$ and ${\bf \Gamma}$. Since there are no restrictions on these parameters, simple random-walk Metropolis steps will be set up, centered at the current value with a component-specific proposal standard deviation chosen so that the acceptance probability is between approximately 0.20 and 0.40. However, since ${\bf \Psi}$ must be a positive definite matrix, a slightly different approach will be taken. The intuition behind this step falls back to the motivation of the model in \cite{hoff_niu}, in which ${\bf \Psi}$ is the covariance matrix of an error term; Hoff suggests using an inverse-Wishart prior distribution. As mentioned previously, an inverse-Wishart prior is undesirable because it is not clear how to specify one in a non-informative manner. However, we can use an inverse-Wishart proposal density for ${\bf \Psi}$, and because ${\bf \Psi}$ represents a baseline kernel matrix in this model, we might center the proposal around an estimate of the geometric anisotropy matrix from the stationary model (with potential scaling to allow the minimum kernel matrix to be smaller than the stationary model). Specifically, an independent Metropolis step will be used with an inverse-Wishart proposal with scale matrix $k \hat{\bf \Sigma}$, where $\hat{\bf \Sigma}$ is the estimated anisotropy matrix from the stationary model and $k$ is a scale factor to control acceptance probability (for this application, we chose $k = 0.65$). The proposal degrees of freedom is set to four, which centers the proposal around $k \hat{\bf \Sigma}$.

The specific ordering of the sampling in the previous step is first ${\bf \Gamma}$, then ${\bf \Psi}$, and finally $\boldsymbol{\alpha}_{-0}$.

\label{lastpage}

\end{document}